\documentclass[preprint,aps,floatfix]{revtex4}
\usepackage{hyperref}
\everymath={\displaystyle}
\usepackage{young}
\usepackage{rotating}
\usepackage{longtable}
\def\1{\mbox{l\hspace{-0.53em}1}}
\newcommand{\fr}{\frac}
\usepackage{multirow}
\newlength{\AccoHaut}

\begin{document}
\title{SU(6) $[{\bf 70},1^-]$ baryon multiplet in the $1/N_c$ expansion}

\author{N. Matagne\footnote{E-mail address: nicolas.matagne@umons.ac.be}}
\affiliation{University of Mons, Service de Physique Nucl\'eaire et 
Subnucl\'eaire,Place du Parc 20, B-7000 Mons, Belgium}

\author{Fl. Stancu\footnote{E-mail address: fstancu@ulg.ac.be}}
\affiliation{University of Li\`ege, Institute of Physics B5, Sart Tilman,
B-4000 Li\`ege 1, Belgium}

\date{\today}

\begin{abstract}
The  masses of excited states of mixed orbital symmetry of
nonstrange  and strange baryons belonging to the lowest  $[{\bf 70},1^-]$ multiplet
are calculated in the $1/N_c$ expansion  to order $1/N_c$ with a new method which
allows to considerably reduce the number of linearly independent operators 
entering the mass formula. This study represents an extension to SU(6) of our work
on nonstrange baryons, the framework of which was SU(4). The conclusion 
regarding the role of the flavor operator, neglected in previous SU(6) studies, is
reinforced.  Namely, both the flavor and spin operators contribute dominantly
to the flavor-spin breaking.
\end{abstract}

\maketitle

%\begin{keyword}
 
%\end{keyword}

\section{Introduction}
Already in 1974 't Hooft \cite{HOOFT} introduced a perturbative expansion
of QCD in terms of the parameter $1/N_c$  where $N_c$ is the number
of colors, to be used in a nonperturbative  QCD regime.  Combined with 
the power counting rules of Witten \cite{WITTEN} 
the   $1/N_c$ expansion method became a powerful tool for a systematic
analysis of baryon properties, both qualitatively and quantitatively.
The success of the method stems from the discovery that the ground state baryons have 
an exact contracted SU(2$N_f$) symmetry 
when $N_c \rightarrow \infty $  \cite{Gervais:1983wq,DM},  $N_f$
being the number of flavors.
For $N_c \rightarrow \infty $ the baryon masses are degenerate. 
For finite $N_c$ the contracted flavor-spin symmetry is broken by effects suppressed by powers of $1/N_c$
so that the mass splitting starts at order $1/N_c$.
As a consequence a considerable amount of work has been devoted to the ground state
baryons  \cite{DM, Jenk1,DJM94,DJM95,CGO94,JL95,DDJM96}.
Operator reduction rules simplify the $1/N_c$ expansion
\cite{DJM94,DJM95} and 
it is customary to drop higher order corrections of order $1/N^2_c$.
Recently   it has been shown that lattice data clearly display both the $1/N_c$ and SU(3) flavor symmetry 
breaking hierarchy \cite{Jenkins:2009wv}.

Under the assumption that 't Hooft's suggestion \cite{HOOFT} would lead to
an $1/N_c$ expansion in all QCD regimes,
the applicability of the approach to excited states has been discussed
since 1994 and it still remains an open problem for deriving the mass spectrum.

>From the  baryon spectrum, the $[{\bf 70},1^-]$ multiplet 
has been most extensively studied. 
For $N_f$ = 2 there are numerous studies as for example Refs.
\cite{CGKM,Goi97,PY1,PY2,CCGL,CaCa98,Pirjol:2003ye,Cohen:2003tb}.
To our knowledge the   $N_f$ = 3 case has been considered  only in Ref.  \cite{SGS}, where
first order corrections in SU(3) symmetry breaking 
were  also included.  In either case, the conclusion was that the splitting starts 
at order $N^0_c$. 

The above studies were based on a method where the system of
$N_c$ quarks was decoupled into a ground state core of $N_c-1$ quarks and an excited 
quark \cite{CCGL}.
This implies  that each generator of SU($2N_f$) has to be written as a sum of two
terms, one acting on the excited quark and the other on the core. 
As a consequence, the number of linearly independent operators $O_i$  in the 
mass formula increases 
tremendously  and the number of the coefficients $c_i$,   encoding the quark dynamics, 
to be determined numerically by a fit,  becomes much larger than the experimental data
available, as for example for the lowest negative parity nonstrange baryons \cite{CCGL}. Accordingly, 
the choice of the most dominant operators in the mass formula becomes 
out of control  which implies 
that important physical effects can be missed, as discussed below.

A technical advantage of the decoupling scheme is to 
reduce the problem of  the knowledge of matrix elements of SU(2N$_f$) generators
between mixed symmetric states $[N_c-1,1]$ to the knowledge of the matrix elements of
these generators
between symmetric states $[N_c-1]$, which are easier to find than those of mixed symmetric 
states (see below).

As an alternative approach, we have recently proposed a new method  \cite{Matagne:2006dj}
where the core + quark separation  is avoided.  Then  we deal with SU(2N$_f$)
generators acting on the whole system.  Accordingly, the number of linearly independent 
operators turns out to be smaller than the number of data.   
All these operators can be included in the fit to clearly find out the most dominant
ones up to order $1/N_c$. The knowledge of matrix elements of SU(2N$_f$)
generators between mixed symmetric states $[N_c-1,1]$ is necessary.

In this approach we have first analyzed the nonstrange $[{\bf 70},1^-]$ multiplet 
where the algebraic work was based on Ref. \cite{HP} which provided the matrix 
elements in terms of isoscalar factors of SU(4), initially derived in the context of
nuclear physics but quite easily applicable to a system of $N_c$ quarks. In this way 
we  have shown that the flavor (in this case the isospin) term becomes as dominant
in $\Delta$ resonances as the spin term in $N$ resonances.   Note that the flavor 
operator was omitted in Ref.  \cite{CCGL}.

Due to this promise we proceeded further with the algebraic work to extend the method to SU(6).
We have first obtained the matrix elements of
all SU(6) generators between symmetric $[N_c]$ states  \cite{Matagne:2006xx} and
in the next step the matrix elements between mixed symmetric states $[N_c-1,1]$ states  \cite{Matagne:2008kb}.
According to the generalized  Wigner-Eckart theorem  described in Ref. \cite{HP}
this amounts at finding the corresponding isoscalar factors.  

Based on the knowledge of these isoscalar factors the present work 
can be seen as an extension of our previous analysis of nonstrange baryons to both
nonstrange and strange baryons,  within the method proposed
in Ref. \cite{Matagne:2006dj}.

The work is organized as follows. 
In the next section we define the SU(6) $\times$ O(3) basis states. In Sec. III we recall  
the SU(6) algebra and introduce the matrix elements of the SU(6) generators by using a generalized Wigner-Eckart
theorem.  The mass operator is described in Sec. IV and  the results of the fit in Sec. V.  Sec. VI 
deals with the conclusion.

In Appendix A we recall some symmetry properties of the isoscalar factors of SU(3) 
and SU(6).  In Appendix B we derive the analytic form of the matrix elements of some operators
entering the mass  formula, relevant for this work.  In Appendix C we give  the full list of
the isoscalar factors associated to the multiplets $^28$ and $^48$ by 
completing in this way the results derived in  Ref. \cite{Matagne:2008kb}.
In Appendix D we introduce two SU(3) breaking operators, give the general formula of the 
matrix elements of the breaking terms in each case and exhibit tables containing the analytic expressions of these 
matrix elements as a function of $N_c$, the strangeness $\mathcal{S}$ and the isospin $I$. 

%%%%%%%%%%%%%%%%%%%%%%%%%%%%%%%%%%%%%%%%%%%%%%%%%%%%%%%%%%%%%%%%%%%%%%%%%%%%%%% 
\section{The wave function}

We deal with a system of $N_c$ quarks having one unit of
orbital excitation. Then the orbital wave function must have a mixed 
symmetry $[N_c-1,1]$. Its spin-flavor part 
must have the same symmetry in order to obtain a totally symmetric
state in the orbital-spin-flavor space. 
The general form of such a wave function is \cite{Stancu:1991rc}
\begin{equation}
\label{EWF}
|[N_c] \rangle = {\frac{1}{\sqrt{d_{[N_c-1,1]}}}}
%|[N_c] \rangle = {\left(\frac{1}{d_{[N_c-1,1]}}\right)}^{1/2}
\sum_{Y} |[N_c-1,1] Y \rangle_{O}  |[N_c-1,1] Y \rangle_{FS}
\end{equation}    
where $d_{[N_c-1,1]} = N_c - 1$ is the dimension of the representation 
$[N_c-1,1]$ of the permutation group $S_{N_c}$ and $Y$ is a symbol for a
Young tableau (Yamanouchi symbol). 
The sum is performed over all possible standard Young tableaux. 
In each term the first basis vector represents the orbital  
space ($O$) and the second the spin-flavor space ($FS$).
In this sum there is only one $Y$ (the normal Young tableau) 
where the last particle is in 
the second row and $N_c - 2 $ terms where the last particle
is in the first row. 

If there is no decoupling, there is no need to
specify $Y$, the matrix elements being identical for all $Y$'s,
due to Weyl's duality between a linear group and a symmetric group
in a given tensor space.
% \footnote{see Ref. \cite{book}, Sec 4.5.}.

Then in SU(6) $\times$ SO(3) the most general form of the wave 
function for a state of a given SU(6) symmetry 
$[f]$ and total angular momentum $J$ and projection $J_3$ is given by
\begin{equation}\label{WF}
|\ell S;JJ_3;(\lambda \mu) Y I I_3\rangle  =%}\nonumber \\
\sum_{m_\ell,S_3}
      \left(\begin{array}{cc|c}
	\ell    &    S   & J   \\
	m_\ell  &    S_3  & J_3 
      \end{array}\right)                
| [f] (\lambda \mu ) YII_3; \ell S J J_3 \rangle
                     |\ell m_\ell \rangle,
\end{equation}
where presently we are interested in $[f] = [N_c-1,1]$.

%%%%%%%%%%%%%%%%%%%%%%%%%%%%%%%%%%%%%%%%%%%%%%%%%%%%%%%%%%%%%%%%%%%%%%%%%%%
\section{SU(6) generators as tensor operators}

We recall that the group SU(6) has 35 generators ${S^i,T^a,G^{ia}}$
with $i = 1,2,3$ and $a = 1,2,\ldots,8$ where $S^i$ are the generators 
of the spin subgroup SU(2) and $T^a$ the generators of the flavor 
subgroup SU(3). The group algebra is
\begin{eqnarray}\label{ALGEBRA}
&[S^i,S^j]  =  i \varepsilon^{ijk} S^k,
~~~~~[T^a,T^b]  =  i f^{abc} T^c, \nonumber \\
&[S^i,G^{ja}]  =  i \varepsilon^{ijk} G^{ka},
~~~~~[T^a,G^{jb}]  =  i f^{abc} G^{ic}, \nonumber \\
&[G^{ia},G^{jb}] = \fr{i}{4} \delta^{ij} f^{abc} T^c
+\fr{i}{2} \varepsilon^{ijk}\left(\fr{1}{3}\delta^{ab} S^k 
+d^{abc} G^{kc}\right),
\end{eqnarray}
by which the normalization of the generators is fixed.

We redefine the generators forming the algebra (\ref{ALGEBRA}) as 
\begin{equation} \label{normes}
E^i =\frac{ S^i}{\sqrt{3}};~~~ E^a = \frac{T^a}{\sqrt{2}}; ~~~E^{ia} = \sqrt{2}
G^{ia} .
\end{equation}
Note that the generic name  for every generator will also be $E^{ia}$ \cite{HP}.
Specifications will be made whenever necessary.
Here we search for the matrix elements of  $S^i$, $T^a$ 
and $G^{ia}$ between SU(6) states of symmetry $[N_c-1,1]$.
As we shall see below, the matrix elements of $S^i$ and $T^a$
are straightforward. The remaining problem is to derive
the matrix elements of $G^{ia}$.

By analogy to SU(4)  \cite{HP} one can write
the  matrix elements of every SU(6) generator $E^{ia}$ as
\begin{eqnarray}\label{GEN}
\lefteqn{\langle [N_c-1,1](\lambda' \mu') Y' I' I'_3 S' S'_3 | E^{ia} |
[N_c-1,1](\lambda \mu) Y I I_3 S S_3 \rangle =}\nonumber \\ & & \sqrt{C^{[N_c-1,1]}(\mathrm{SU(6)})} %  \nonumber \\
   \left(\begin{array}{cc|c}
	    S   &    S^i   & S'   \\
	    S_3  &   S^i_3   & S'_3
  \end{array}\right)
     \left(\begin{array}{cc|c}
	I   &   I^a   & I'   \\
	I_3 &   I^a_{3}   & I'_3
   \end{array}\right)  \nonumber \\
& & \times       \sum_{\rho = 1,2}
 \left(\begin{array}{cc||c}
	(\lambda \mu)    &  (\lambda^a\mu^a)   &   (\lambda' \mu')\\
	Y I   &  Y^a I^a  &  Y' I'
      \end{array}\right)_{\rho}
\left(\begin{array}{cc||c}
	[N_c-1,1]    &  [21^4]   & [N_c-1,1]   \\
	(\lambda \mu) S  &  (\lambda^a\mu^a) S^i  &  (\lambda' \mu') S'
      \end{array}\right)_{\rho} , %\nonumber \\
   \end{eqnarray}
where $C^{[N_c-1,1]}(\mathrm{SU(6)}) = N_c(5 N_c+18)/12$ is the SU(6)
Casimir operator associated to 
the irreducible representation $[N_c-1,1]$, followed by Clebsch-Gordan
coefficients of SU(2)-spin and SU(2)-isospin. The sum over $\rho$ is over
terms containing products of isoscalar factors of SU(3) and SU(6) respectively.
In particular, $T^a$ is an SU(3) irreducible tensor operator of components
$T^{(11)}_{Y^aI^a}$, %\emph{i.e.}
where $a$ corresponds to $(\lambda^a\mu^a) = (11) $ in the present case. 
It is a scalar in SU(2) so that the index $i$ from $E^{ia}$
is no more necessary. The generators $S^i$
form a rank 1 tensor in SU(2) which is a scalar in SU(3), so the index $i$ suffices.
Although we use the same symbol for the operator $S^i$ and its quantum numbers 
we hope that no confusion is created.
Thus, for the generators $S^i$ and $T^a$, which are elements of the $su$(2) and
$su$(3) subalgebras of (\ref{ALGEBRA}), the above expression
simplifies considerably. In particular, as  $S^i$  acts only on the
spin part of the wave function, we apply the usual
Wigner-Eckart theorem for SU(2) to get
\begin{eqnarray}\label{SPIN}
\langle [N_c-1,1](\lambda'\mu') Y' I' I'_3; S' S'_3 |S^i|
[N_c-1,1](\lambda \mu) Y I I_3; S S_3 \rangle = \nonumber \\   \delta_{SS'}\delta_{\lambda \lambda'} \delta_{\mu\mu'} \delta_{YY'} \delta_{II'} \delta_{I_3I_3'}%\nonumber \\
 \sqrt{C(\mathrm{SU(2)})} \left(\begin{array}{cc|c}
	S   &    1  &  S'   \\
	S_3 &    i  &  S'_3
      \end{array}\right),
   \end{eqnarray}
with $C(\mathrm{SU(2)}) = S(S+1)$.  
Similarly, we use the Wigner-Eckart theorem for $T^a$ which is a generator of 
the subgroup SU(3), so we have
\begin{eqnarray}\label{FLAVOR}
\lefteqn{\langle [N_c-1,1](\lambda'\mu') Y' I' I'_3; S' S'_3 |T_a|
[N_c-1,1](\lambda \mu) Y I I_3; S S_3 \rangle =} \nonumber \\ & &
\delta_{SS'} \delta_{S_3S'_3}\delta_{\lambda \lambda'} \delta_{\mu\mu'}
%\nonumber \\
%\times
\sum_{\rho = 1,2}
\langle (\lambda'\mu') || T^{(11)} || (\lambda \mu) \rangle_{\rho}
  \left(\begin{array}{cc|c}
	(\lambda \mu)    &  (11)   &   (\lambda'\mu')\\
	YII_3   &  Y^aI^aI^a_{3}  &  Y' I' I'_3
      \end{array}\right)_{\rho},
   \end{eqnarray}
where the reduced matrix element is defined as  \cite{HECHT} 
\begin{eqnarray}\label{REDUCED}
\langle (\lambda \mu) || T^{(11)} || (\lambda \mu) \rangle_{\rho} = \left\{
\begin{array}{cc}
\sqrt{C(\mathrm{SU(3)})}      & \mathrm{for}\ \rho = 1 \\
0 & \mathrm{for}\ \rho = 2 \\
\end{array}\right.
,\end{eqnarray}   
in terms of the eigenvalue of the Casimir operator       
$C(\mathrm{SU(3)}) = \frac{1}{3} g_{\lambda \mu}$ where
\begin{equation}\label{CSU3}
g_{\lambda\mu}= {\lambda}^2+{\mu}^2+\lambda\mu+3\lambda+3\mu.
\end{equation}
Note that the presence of the index $\rho$ has the same origin as in Eq. 
(\ref{GEN}), namely it reflects the 
multiplicity problem appearing in the direct product of SU(3) irreducible representations 
\begin{eqnarray}\label{PROD}
\lefteqn{(\lambda \mu) \times (11)  =  (\lambda+1, \mu+1)+ (\lambda+2, \mu-1) +
(\lambda \mu)_1 + (\lambda \mu)_2}  \nonumber \\
& & + \, (\lambda-1, \mu+2) + (\lambda-2, \mu+1)
+ (\lambda+1, \mu-2)+ (\lambda-1, \mu-1).
\end{eqnarray}
%But according to the definition Eq. (\ref{REDUCED}) only $\rho = 1$ contributes.
 Each SU(3) CG coefficient factorizes 
into an SU(2)-isospin CG coefficient and an SU(3) isoscalar factor
\cite{DESWART}
\begin{equation}\label{CGSU3}
\left(\begin{array}{cc|c}
	(\lambda \mu)    &  (11)   &   (\lambda'\mu')\\
	YII_3   &  Y^aI^aI^a_{3}  &  Y' I' I'_3
      \end{array}\right)_{\rho} =
\left(\begin{array}{cc|c}
	I   &    1  &  I'   \\
	I_3 &    I^a_3  &  I'_3
     \end{array}\right)
 \left(\begin{array}{cc||c}
	(\lambda \mu)    &  (11)   &   (\lambda'\mu')\\
 	 YI   &  Y^aI^a  &  Y' I'
      \end{array}\right)_{\rho}.
 \end{equation}   
%The $\rho$ dependence is consistent with Eq. (\ref{GEN}).
The analytic expression of the isoscalar factors
can be found in Table 4 of Ref. \cite{HECHT}.

%%%%%%%%%%%%%%%%%%%%%%%%%%%%%%%%%%%%%%%%%%%%%%%%%%%%%%%%%%

%%%%%%%%%%%%%%%%%%%%%%%%%%%%%%%%%%%%%%%%%%%%%%%%%%%%%%%%%%%%%%%%%%%%
\section{The mass operator}

When SU(3) is broken the mass operator takes the following general 
form  as first proposed in Ref. \cite{JL95} for the  symmetric baryon multiplet $[N_c]$
\begin{equation}
\label{massoperator}
M = \sum_{i}c_i O_i + \sum_{i}d_i B_i .
\end{equation} 
The operators $O_i$ are of type (\ref{OLFS})
\begin{equation}\label{OLFS}
O_i = \frac{1}{N^{n-1}_c} O^{(k)}_{\ell} \cdot O^{(k)}_{SF},
\end{equation}
where  $O^{(k)}_{\ell}$ is a $k$-rank tensor in SO(3) and  $O^{(k)}_{SF}$
a $k$-rank tensor in SU(2)-spin, but invariant in SU($N_f$).
Thus $O_i$ are rotational invariant.
For the ground state one has $k = 0$. The excited
states also require  $k = 1$ and $k = 2$ terms.

The rank $k = 2$ tensor operator of SO(3) is
\begin{equation}\label{TENSOR} 
L^{(2)ij} = \frac{1}{2}\left\{L^i,L^j\right\}-\frac{1}{3}
\delta_{i,-j}\vec{L}\cdot\vec{L},
\end{equation}
which, like $L^i$, acts on the orbital wave function $|\ell m_{\ell} \rangle$  
of the whole system of $N_c$ quarks (see  Ref. \cite{Matagne:2005gd} for the normalization 
of $L^{(2)ij}$). 

The operators  $B_i$ are 
SU(6) breaking and are defined to have zero expectation values for nonstrange baryons.
The values of the coefficients $c_i$ and $d_i$
which encode the QCD dynamics are determined from numerical fits to data.
They are presented below.

Table \ref{ops} gives the list of all  linearly independent operators of type (\ref{OLFS})
organized by  powers  of $1/N_c$ in their matrix  ($N_c$,  $N^0_c$ and $N^{-1}_c$). Our choice of operators entering the mass 
formula (\ref{massoperator}) is presented in
Table  \ref{operators},  which contains the most relevant  $O_i$  
of Table \ref{ops}. Unlike to SU(4) (two flavors), in the SU(6) case (three flavors),
the $N_c$ order of the matrix elements of a given operator $O_i$
is not always the same for all multiplets, as one can see from Table \ref{Matrix}.

\begin{table}
\caption{The  linearly independent spin-singlet flavor-singlet
operators for $N_f=3$, organized by powers of $1/N_c$ in their matrix
elements.  \label{ops}}\renewcommand{\arraystretch}{2.3} {\scriptsize
\begin{tabular}{c|c}
Order of matrix element &
Operator \\ \hline\hline

$N_c$ & $N_c\ \1$ \\ \hline

$N_c^0$ & $\ell\cdot s,\  \frac{1}{N_c} \left[T \cdot T - \frac{1}{12} N_c(N_c+6)\right], \ \frac{3}{N_c} L\cdot T \cdot G, \ 
  \frac{15}{N_c} L^{(2)}\cdot G\cdot G, \ \frac{1}{N_c^2}L\cdot G\cdot \{ S, G\}$ \\ \hline
$N_c^{-1}$ & $\frac{1}{N_c} S\cdot S, \
\frac{1}{N_c} L^{(2)}\cdot S \cdot S, \ 
\frac{1}{N_c^2} L^{(2)}\cdot T \cdot\{ S, G\},\ \frac{3}{N_c^2} S\cdot T\cdot G $ \\ \hline

\end{tabular}}
\end{table}

 As far as the SU(6) breaking is concerned we have first selected  the most
dominant operator $B_1 = \mathcal{S}$, where  $\mathcal{S}$ is the strangeness. Next we have 
introduced an  operator  named $B_2$, which was found to play an important role, as discussed below. 

\begin{table*}[h!]%resp2_ed
\caption{Operators and their coefficients in the mass formula obtained from 
numerical fits. The values of $c_i$ and $d_i$ are indicated under the heading Fit $n\ (n=1,2,3)$,
in each case.}
\label{operators}{\scriptsize
\renewcommand{\arraystretch}{2} % enlarge line spacing
\begin{tabular}{lrrr}
\hline
\hline
Operator \hspace{2cm} &\hspace{0.0cm} Fit 1 (MeV) & \hspace{0.cm} Fit 2 (MeV) & Fit 3 (MeV)  \\
\hline
%\hline%
$O_1 = N_c \ \1 $                    & $489 \pm 4$  & $492 \pm 4$  & $492\pm 4$       \\
$O_2 = \ell^i s^i$                	     & $24 \pm 6$ & $6 \pm 6$  & $6\pm5$   \\
$O_3 = \frac{1}{N_c}S^iS^i$          & $129 \pm 10$ & $123 \pm 10$ & $123\pm 10$   \\
$O_4 = \frac{1}{N_c} \left[T^aT^a - \frac{1}{12} N_c(N_c+6)\right]$  & $145 \pm 16$ & $134\pm 16$ & $135\pm 16$    \\
$O_5 =  \frac{3}{N_c} L^i T^a G^{ia}$ & $-19 \pm 7$ & $3 \pm 7$ & $4\pm 3$   \\ 
$O_6 = \frac{15}{N_c} L^{(2)ij}G^{ia}G^{ja}$    & $9 \pm 1$ & $9\pm 1$ & $9\pm 1$        \\
$O_7 = \frac{1}{N_c^2}L^iG^{ja}\{S^j,G^{ia}\}$ & $129 \pm 33$ &  $6\pm 33$  &\\ 
\hline
$B_1 = \mathcal{-S}$  & $138 \pm 8$ & $138\pm 8$ & $137\pm 8$\\
%$B_2 = \frac{1}{N_c} \left(T^3T^3 - O_4\right)$ &  $-59\pm 18$ & $-40\pm 18$ & $-40\pm 18$\\ 
$B_2 = \frac{1}{N_c} \sum^3_{i=1}T^iT^i - O_4$ &  $-59\pm 18$ & $-40\pm 18$ & $-40\pm 18$\\ 
\hline
$\chi_{\mathrm{dof}}^2$   &  $1.7$  & $0.9$ & $0.84$   \\
\hline \hline
\end{tabular}}
\end{table*}

\begin{table*}[h!]%[htb]
\begin{center}
\caption{Matrix elements of $O_i$ for all states belonging to the 
$[{\bf 70},1^-]$ multiplet. The vanishing off-diagonal matrix elements are not included.  }
\label{Matrix}
\renewcommand{\arraystretch}{2.3} {\scriptsize
\begin{tabular}{lccccccccccc}
\hline
\hline
  &\hspace{0cm}  $O_1$ &\hspace{0cm} $O_2$ & \hspace{0cm}$O_3$ &&\hspace{0cm}
  $O_4$ & \hspace{0cm}$O_5$ && \hspace{0cm}$O_6$    &  $O_7$ &\hspace{0cm}     \\
  \hline
$^28_{\frac{1}{2}}$ & $N_c$  & $-\frac{2N_c-3}{3N_c}$ & $\frac{3}{4N_c}$ 
&& $\frac{3}{4N_c}$     &  $ -\frac{3}{N_c} $ 
 & &   $0$ & $\frac{5-N_c(3N_c+10)}{24N_c^2}$ &\\
%%%%%%%%%%%%%%%%%%%%%%%%%%%%%%%%%%%%%%%%%%%%%%%%%%%%%%%%%%%%%%%%%%%%%%%%%%%%%%
$^48_{\frac{1}{2}}$ & $N_c$   &  $-\frac{5}{6}$ & $\frac{15}{4N_c}$ 
&& $\frac{3}{4Nc}$     & $-\frac{5(N_c+3)}{4N_c}$  
 & & $-\frac{25(N_c-1)}{8N_c}$ &  $-\frac{5(N_c-1)(3N_c+5)}{48N_c^2}$ & \\
%%%%%%%%%%%%%%%%%%%%%%%%%%%%%%%%%%%%%%%%%%%%%%%%%%%%%%%%%%%%%%%%%%%%%%%%%%%%
$^28_{\frac{3}{2}}$ & $N_c$  & $\frac{2N_c-3}{6N_c}$ & $\frac{3}{4N_c}$  
&& $\frac{3}{4N_c}$     &$ \frac{3}{2N_c}$  
 & & $0$  & $-\frac{5-N_c(3N_c+10)}{48N_c^2}$ &\\
%%%%%%%%%%%%%%%%%%%%%%%%%%%%%%%%%%%%%%%%%%%%%%%%%%%%%%%%%%%%%%%%%%%%%%%%%%%%
$^48_{\frac{3}{2}}$ & $N_c$  &  $-\frac{1}{3}$ & $\frac{15}{4N_c}$ 
&& $\frac{3}{4N_c}$ &    $ -\frac{N_c+3}{2N_c} $  
 & & $\frac{5(N_c-1)}{2N_c}$  & $-\frac{(N_c-1)(3N_c+5)}{24N_c^2}$ &\\
%%%%%%%%%%%%%%%%%%%%%%%%%%%%%%%%%%%%%%%%%%%%%%%%%%%%%%%%%%%%%%%%%%%%%%%%%%
$^48_{\frac{5}{2}}$ & $N_c$ & $\frac{1}{2}$ & $\frac{15}{4N_c}$ &&
$\frac{3}{4N_c}$ &    $ \frac{3(N_c+3)}{4N_c} $  
 & & $-\frac{5(N_c-1)}{8N_c}$ & $\frac{(N_c-1)(3N_c+5)}{16N_c^2}$ &\\
%%%%%%%%%%%%%%%%%%%%%%%%%%%%%%%%%%%%%%%%%%%%%%%%%%%%%%%%%%%%%%%%%%%%%%%%%
$^2{10}_{\frac{1}{2}}$ & $N_c$ & $\frac{1}{3}$ & $\frac{3}{4N_c}$ &&
$\frac{15}{4N_c}$ &    $-\frac{3(N_c+1)}{2N_c} $   
 & & $0$ & $\frac{17 - N_c (3 N_c+10)}{24 N_c^2}$ & 
 \\
%%%%%%%%%%%%%%%%%%%%%%%%%%%%%%%%%%%%%%%%%%%%%%%%%%%%%%%%%%%%%%%%%%%%%%%
$^2{10}_{\frac{3}{2}}$ & $N_c$ & $-\frac{1}{6}$ & $\frac{3}{4N_c}$ &&
$\frac{15}{4N_c}$ &    $  \frac{3(N_c+1)}{4N_c}$   
 & & $0$ 
& $-\frac{17 - N_c (3 N_c+10)}{48 N_c^2}$ &\\
%%%%%%%%%%%%%%%%%%%%%%%%%%%%%%%%%%%%%%%%%%%%%%%%%%%%%%%%%%%%%%%%%%
$^2{1}_{\frac{1}{2}}$ & $N_c$ & $-1$ & $\frac{3}{4N_c}$ &&
$ -\frac{2N_c+3}{4N_c} $ &     $\frac{N_c-3}{2N_c}$  
 & & $0$ & $-\frac{1 + 3 N_c (N_c+4)}{24 N_c^2}$ & \\
%%%%%%%%%%%%%%%%%%%%%%%%%%%%%%%%%%%%%%%%%%%%%%%%%%%%%%%%%%%%%%%%%%
$^2{1}_{\frac{3}{2}}$ & $N_c$ & $\frac{1}{2}$ & $\frac{3}{4N_c}$ &&
$ -\frac{2N_c+3}{4N_c}$ &     $-\frac{N_c-3}{4N_c}$   
 & & $0$ & $\frac{1 + 3 N_c (N_c+4)}{48 N_c^2}$ & \\
%%%%%%%%%%%%%%%%%%%%%%%%%%%%%%%%%%%%%%%%%%%%%%%%%%%%%%%%%%%%%%%%%%%%
$^28_{\frac{1}{2}} -$ $^48_{\frac{1}{2}}$ & 0 & $-\frac{1}{3}\sqrt{\frac{N_c+3}{2N_c}}$ & 0 && 0  &
$\frac{1}{2N_c}\sqrt{\frac{N_c(N_c+3)}{2}}$ &&
$-\frac{25}{4N_c}\sqrt{\frac{N_c(N_c+3)}{2}}$ & $\frac{5}{12N_c^2}\sqrt{\frac{N_c(N_c+3)}{2}} $ & \\
$^28_{\frac{3}{2}}-$ $^48_{\frac{3}{2}}$ & 0 & $-\frac{1}{6}\sqrt{\frac{5(N_c+3)}{N_c}}$ & 0 && 0  &
 $\frac{1}{4N_c}\sqrt{5N_c(N_c+3)}$ &&
$\frac{5}{8N_c}\sqrt{5N_c(N_c+3)}$ & $\frac{5}{24N_c^2}\sqrt{5N_c(N_c+3)}$ &
\vspace{0.15cm} \\
%& & & & & & \\
\hline \hline
\end{tabular}}
\end{center}
\end{table*}

%%%%%%%%%%%%%%%%%%%%%%%%%%%%%%%%%%%%%%%%%%%%%%%%%%%%%%%%%%%%%%%%%%%%%%%%%%%%%%%%%%

In Table   \ref{operators} 
the first nontrivial operator is the spin-orbit operator $O_2$.
In the spirit of the Hartree picture \cite{WITTEN}, generally adopted 
for the description of baryons,  we identify the 
spin-orbit operator with the single-particle operator 
\begin{equation}\label{spinorbit}
\ell \cdot s = \sum^{N_c}_{i=1} \ell(i) \cdot s(i),
\end{equation}
the matrix elements of which are of order $N^0_c$.
For simplicity  we ignore 
the two-body part of the spin-orbit operator, denoted by
$1/N_c\left(\ell \cdot S_c\right)$ in Ref. \cite{CCGL},
as being of a lower order
(the lower case indicates operators acting on
the excited quark and the subscript $c$ is attached to those
acting on the core).
The spin operator $O_3$ and the flavor operator $O_4$ are two-body and linearly independent. 
The operators  $O_5$ and $O_6$,  are  two-body, which means that they carry a 
factor $1/N_c$.  But as $G^{ia}$
sums coherently, it introduces an extra factor $N_c$ and makes the matrix 
elements of $O_5$  and $O_6$ of order $N^0_c$  as well, except for $O_5$
in the $^28$ multiplet.

The operators $O_5$ and $O_6$ are normalized to allow their 
coefficients $c_i$ to have a natural size \cite{SGS,Scoccola:2007sn}. 
The normalization factors result from the matrix elements 
of $O_i$ presented in Table \ref{Matrix}.
We have also included the more complex operator $O_7$, which contains an
anticommutator and is three-body.  But because it contains the coherent generator $G^{ia}$ 
two times, its matrix elements turn out to be of order $\mathcal{O}(N^0_c)$. The presence of 
coherent factors in a composite operator leads to an enhancement of the 
order of the matrix elements, as already known from the SU(4) case  \cite{CCGL}.

The matrix elements of the operators $O_i$ have been calculated for all available states of the 
multiplet $[{\bf 70},1^-]$ starting from the wave function (\ref{WF}) and 
using the isoscalar factors of Ref.  \cite{Matagne:2008kb} combined with Tables
\ref{octet_spin_one_half},   \ref{octet_spin_three_halfs}, 
\ref{decuplet_spin_one_half} 
 and  \ref{singlet_spin_one_half} 
of the present work. For completeness 
the general analytic expressions of $O_5$, $O_6$ and $O_7$, 
are given in Appendix B,  up to an obvious factor.
In Table \ref{Matrix} the nonvanishing off-diagonal matrix elements are also indicated whenever the case.  

One would think of also including  the operator 
\begin{equation}\label{tensorspin}
O_8 = \frac{1}{N_c} L^{(2)ij} S^i  S^j,
\end{equation}  
of order $1/N_c$.  
However, interestingly, in our basis
we found a proportionality relation between the following expectation values 
\begin{equation}
\langle L^{(2)ij} S^i  S^j \rangle = 
- \frac{12}{N_c - 1} \langle L^{(2)ij} G^{ia} G^{ia}\rangle, 
\end{equation}  
for all states belonging to the $[{\bf 70},1^-]$ multiplet.
This implies that we cannot include $O_8$  independently  
in the fit to the experimental spectrum, because its expectation 
values are proportional to those of $O_6$ and we ignore the off-diagonal 
matrix elements of $\langle O_8 \rangle$.

The operator $B_2$ was introduced in order to obtain a splitting between $\Sigma$
and $\Lambda$ baryons, like in the
Gell-Mann - Okubo mass formula. Its matrix elements  are given by 
\begin{equation}
\label{B2}
B_2=\frac{1}{N_c} I(I+1)-O_4
\end{equation}
where $O_4$ can be found in column 5 of Table \ref{Matrix}.

There are two other SU(3) breaking operators defined in Appendix D together with their  
matrix elements  given in Tables  \ref{Matrixsg} and  \ref{Matrixlg}.
We found that their contribution is negligible in improving the
fit. An account of such findings can be found in Ref. \cite{BLED}.
%%%%%%%%%%%%%%%%%%%%%%%%%%%%%%%%%%%%%%%%%%%%%%%%%%%%%%%%%%%%%%%%%%%%%%%%%%%%%%%%%%%%%%%%%

{\squeezetable
\begin{table}
%\begin{sidewaystable}
\caption{The partial contribution and the total mass (MeV) predicted by the $1/N_c$ expansion
obtained from the Fit 1.  The last two columns give  the empirically known masses  {\protect \cite{PDG}} and the resonance name and status.}\label{MASSES}
\renewcommand{\arraystretch}{1.0}
\begin{tabular}{crrrrrrrrrcccl}\hline \hline
                    &      \multicolumn{9}{c}{Part. contrib. (MeV)}   & \hspace{.0cm} Total (MeV)   & \hspace{.0cm}  Exp. (MeV)\hspace{0.0cm}& &\hspace{0.cm}  Name, status \hspace{.0cm} \\

\cline{2-10}
                    &   \hspace{.0cm}   $c_1O_1$  & \hspace{.0cm}  $c_2O_2$ & \hspace{.0cm}$c_3O_3$ & $c_4O_4$ &\hspace{.0cm}  $c_5O_5$ &\hspace{.0cm}  $c_6O_6$ & $c_7O_7$ &  $d_1B_1$ & $d_2B_2$&     &        \\
\hline
$N_{\frac{1}{2}}$        & 1467 & -8 &  32 & 36 & 19  & 0 & -31 & 0   & 0 &  $1499\pm 10$ & $1538\pm 18$ & & $S_{11}(1535)$****  \\
$\Lambda_{\frac{1}{2}}$  &      &    &     &    &     &   &     & 138 & 15&  $1668\pm 9$  & $1670\pm 10$  & & $S_{01}(1670)$**** \\
$\Sigma_{\frac{1}{2}}$   &      &    &     &    &     &   &     & 138 &-25&  $1628\pm 10$ &              & & \\
$\Xi_{\frac{1}{2}}$      &      &    &     &    &     &   &     & 276 & 0 &  $1791\pm 13$ &              & & \vspace{0.2cm}\\
\hline
$N_{\frac{3}{2}}$        & 1467 & 4  & 32  & 36 & -10 & 0 & 16  &  0  & 0 &  $1542\pm 10$ & $1523\pm 8$ & & $D_{13}(1520)$****  \\
$\Lambda_{\frac{3}{2}}$  &      &    &     &    &     &   &     & 138 & 15&  $1698\pm  8$ & $1690\pm 5$ & & $D_{03}(1690)$**** \\
$\Sigma_{\frac{3}{2}}$   &      &    &     &    &     &   &     & 138 &-25&  $1658\pm  9$ & $1675\pm 10$             & & $D_{13}(1670)$****\\
$\Xi_{\frac{3}{2}}$      &      &    &     &    &     &   &     & 276 & 0 &  $1821\pm 11$ & $1823\pm 5$             & & $D_{13}(1820)$***
\vspace{0.2cm} \\
\hline
$N'_{\frac{1}{2}}$       & 1467 &-20 &162  & 36 & 48  &-18& 42  & 0   & 0 &  $1648\pm 11$ & $1660\pm 20$ & & $S_{11}(1650)$****  \\
$\Lambda'_{\frac{1}{2}}$ &      &    &     &    &     &   &     & 138 &15 &  $1784\pm 16$ & $1785\pm 65$  & & $S_{01}(1800)$*** \\
$\Sigma'_{\frac{1}{2}}$  &      &    &     &    &     &   &     & 138 &-25&  $1745\pm 17$ & $1765\pm 35$             & & $S_{11}(1750)$***\\
$\Xi'_{\frac{1}{2}}$     &      &    &     &    &     &   &     & 276 & 0 &  $1907\pm 20$ &              & & \vspace{0.2cm}\\
\hline
$N'_{\frac{3}{2}}$       & 1467 & -8 & 162 & 36 & 19  & 15&-17  & 0   & 0 &  $1675\pm 10$ & $1700\pm 50$ & & $D_{13}(1700)$***  \\
$\Lambda'_{\frac{3}{2}}$ &      &    &     &    &     &   &     & 138 &15 &  $1826\pm 12$ &  & &  \\
$\Sigma'_{\frac{3}{2}}$  &      &    &     &    &     &   &     & 138 &-25&  $1787\pm 13$ &              & & \\
$\Xi'_{\frac{3}{2}}$     &      &    &     &    &     &   &     & 276 & 0 &  $1949\pm 16$ &              & & \vspace{0.2cm}\\
\hline
$N_{\frac{5}{2}}$       & 1467 & 12 & 162 & 36 &-29  & -4& 25  & 0   & 0 &  $1669\pm 10$ & $1678\pm 8$ & & $D_{15}(1675)$****  \\
$\Lambda_{\frac{5}{2}}$ &      &    &     &    &     &   &     & 138 & 15&  $1822\pm 10$ & $1820\pm 10$  & & $D_{05}(1830)$**** \\
$\Sigma_{\frac{5}{2}}$  &      &    &     &    &     &   &     & 138 &-25&  $1782\pm 11$ & $1775\pm 5$             & &$D_{15}(1775)$**** \\
$\Xi_{\frac{5}{2}}$     &      &    &     &    &     &   &     & 276 & 0 &  $1945\pm 14$ &              & & \vspace{0.2cm}\\
\hline
$\Delta_{\frac{1}{2}}$   & 1467 & 8  & 32  &181 & 38  & 0 & -24 & 0   & 0 &  $1702\pm 18$  & $1645\pm 30$ & & $S_{31}(1620)$****  \\
$\Sigma''_{\frac{1}{2}}$ &      &    &     &    &     &   &     & 138 & 34&  $1875\pm 16$  &   & &  \\
$\Xi''_{\frac{1}{2}}$    &      &    &     &    &     &   &     & 276 & 59&  $2037\pm 22$  &              & & \\
$\Omega_{\frac{1}{2}}$   &      &    &     &    &     &   &     & 413 & 74&  $2190\pm 29$  &              & & \vspace{0.2cm}\\
\hline
$\Delta_{\frac{3}{2}}$   & 1467 & -4 & 32  & 181& -19 & 0 & 12  &  0  &  0&  $1668\pm 20$  & $1720\pm 50$ & & $D_{33}(1700)$****  \\
$\Sigma''_{\frac{3}{2}}$ &      &    &     &    &     &   &     & 138 & 34&  $1841\pm 16$  &  & &  \\
$\Xi''_{\frac{3}{2}}$    &      &    &     &    &     &   &     & 276 & 59&  $2003\pm 21$  &              & & \\
$\Omega_{\frac{3}{2}}$   &      &    &     &    &     &   &     & 413 & 74&  $2156\pm 27$  &              & & \vspace{0.2cm}\\
\hline
$\Lambda''_{\frac{1}{2}}$&1467  &-24 & 32 &-108 & 0   & 0 &  -38&138  &-44&  $1421\pm 14$  & $1407\pm 4$  & & $S_{01}(1405)$**** \\
\hline
$\Lambda''_{\frac{3}{2}}$&1467  & 12 & 32 &-108 & 0   &  0&  19 & 138 &-44&  $1515\pm 14$    & $1520\pm 1$  & & $D_{03}(1520)$**** \\
\hline
$N_{1/2}-N'_{1/2}$       &0     & -8 &  0 & 0   & -10 &-55& 18  &  0  &  0&  $-55$         &   & & \\
$N_{3/2}-N'_{3/2}$       &0     & -12&  0 & 0   & -15 & 17& 28  &  0  &  0&  18            &  &  & \\
\hline
\hline
\end{tabular}
%\end{sidewaystable}
\end{table}}

%%%%%%%%%%%%%%%%%%%%%%%%%%%%%%%%%%%%%%%%%%%%%%%%%%%%%%%%%%%%%%%
\section{Results}

We have performed three distinct fits of the mass formula (\ref{massoperator}) using the experimental masses from
PDG \cite{PDG}.  There are 17 resonances available, with a status of three or four stars and two mixing angles. 
The latter are defined by the following equations
\begin{eqnarray}
|N^{'}_J \rangle = \cos \theta_J |^4N_J \rangle +
 \sin \theta_J |^2N_J \rangle, \nonumber \\
|N_J \rangle = \cos \theta_J |^2N_J \rangle -
 \sin \theta_J |^4N_J \rangle. 
\end{eqnarray}
Experimentally one finds 
$\theta^{exp}_{1/2} \approx   - 0.56$ rad and
$\theta^{exp}_{3/2} \approx  0.10$ rad  \cite{Hey:1974nc}.

The resulting dynamical coefficients $c_i$ and $d_i$  and the values of $\chi_{\mathrm{dof}}^2$  
are indicated in Table \ref{operators} in each case.
Actually only the Fit 1 is based on the experimental value of  $M(\Lambda(1405)) = 1407$ MeV.  The corresponding 
$\chi_{\mathrm{dof}}^2$ is $1.7$.  
We have tried to understand the somewhat large $\chi_{\mathrm{dof}}^2$  and found out
that it is rather difficult to accommodate the experimentally low mass of $\Lambda(1405)$. 
The situation is entirely similar to all types of quark models, as for example,
\cite{Capstick:1986bm,Glozman:1997ag,Melde:2008yr}. The experimental mixing matrix of the $\Lambda(S01)$ resonances 
in terms of the flavor singlet $^21$ and $^28$ and  $^48$ components \cite{Hey:1974nc}
could not help in lowering  the mass of  $\Lambda(1405)$  \cite{Matagne:2010se}.

To see indeed  that the experimentally low mass of $\Lambda(1405)$  is responsible for the  $\chi_{\mathrm{dof}}^2$
of the Fit 1, we have performed the
Fits 2 and  3  with an arbitrarily  larger value than the experimental mass. We took $M(\Lambda(1405)) = 1500$ MeV,
inspired by quark model calculations.
With this value the  $\chi_{\mathrm{dof}}^2$ goes down to about $0.9$ for these latter fits.   By making the Fit 3 we
explored the role of the operator $O_7$. Removing it from the mass formula the result remained practically unchanged.

However we have to point out the important role of the operator $O_7$ in the Fit 1. Without it the $\chi_{\mathrm{dof}}^2$ 
is about $2.95$, while including, it goes down to $1.7$. 
Actually the operator $O_7$ contains the operators
$O_{9}$ and $O_{11}$ of Ref.  \cite{SGS} and similarly it plays a role in the $\Lambda(1520)$ -  $\Lambda(1405)$ splitting,
enhancing the effect of the spin-orbit operator $O_2$ and leading to  a splitting quite close to the experiment.

As a common feature with the SU(4) case we found  that the isospin operator $O_4$ contributes to the mass 
with a coefficient
$c_4$ very close to that of the spin operators $O_3$.

In Table \ref{MASSES} we present the partial contributions $c_i O_i$ and $d_i B_i$ for all operator included in the mass
formula  (\ref{massoperator}) together with the total mass (MeV)   predicted by the $1/N_c$ expansion in SU(6). 

Similar to  the SU(4) case \cite{Matagne:2006dj} we found  that 
spin operator $O_3$  is dominant in $N$ resonances while  
the flavor operator $O_4$ is dominant in $\Delta$ resonances, with an even larger positive contribution.

This implies that flavor operator is as important as the 
spin operator, a result consistent with that obtained for nonstrange baryons. Thus
these two operators bring the basic contribution to the spin-flavor breaking.  
Note that the operator $O_4$ is also dominant in the flavor singlets $\Lambda''_{\frac{1}{2}}$
and $\Lambda''_{\frac{3}{2}}$. Its negative contribution compensates to a large extent the positive contribution 
of the SU(3) flavor breaking operator $B_1$.  Together with the operator $B_2$, used here for the 
first time, we obtain  the required splitting between $\Lambda$ and $\Sigma$ resonances in a given multiplet.

The terms containing the angular 
momentum components, $O_2$, $O_5$ and $O_6$ are dynamically suppressed,  as suggested by the very small 
values of their coefficients,  $c_2$,   $c_5$ and   $c_6$ respectively.
Although the coefficient $c_7$ is large the contribution $c_7 O_7$ to the mass
is always moderate as one can see from Table \ref{MASSES}.

The flavor breaking operator $B_1$ is important to all strange baryons.

In Ref.  \cite{SGS}  $\Lambda$(1405) acquired a mass very close to experiment.  Our result for
this resonance, $1421 \pm 14$ MeV  is not far from the experimental interval.
There is however some  difference between the dynamics of our approach and that of Ref. \cite{SGS}.  
In Ref. \cite {SGS} only the wave function component with 
$S_c$ = 0 is taken into account and this component brings no contribution to the
spin term in flavor singlets, which means that  the operators $\frac{1}{N_c} S^c \cdot S^c$ and 
$\frac{1}{N_c} s \cdot S^c$ have zero expectation values.
For this basic reason the mass of  $\Lambda (1405)$ 
is not affected by the spin contribution and remains low, while the other masses are moved upwards.
In our case, where we use 
the exact wave function, both  $S_c = 0$ and  $S_c = 1$ parts of the wave function 
contribute to the spin term. This makes the 
spin term expectation value identical for all states of a given $J$  ($\langle O_3 \rangle = 3/(4N_c)$ for $J = 1/2$
and   $\langle O_3 \rangle = 15/(4N_c)$ for $J = 3/2$, see  Table 
\ref{Matrix})
irrespective of the flavor, which seems to
us natural.  Then, in our case,  with a non vanishing spin term in flavour singlets as well,
the mass formula accommodates a heavier  $\Lambda (1405)$ than the experiment,
like in quark models, as mentioned above (for a review on the controversial nature of $\Lambda (1405)$ see, for example,
Ref. \cite{Pakvasa:1999zv}).  
% where one of the authors S.F. Tuan has predicted together with 
%D.H. Dalitz this resonance in 1959, discovered experimentally two years later.)
Interestingly, the isospin operator, absent in Ref.   \cite{SGS}, although of order $\mathcal{O}(N^0_c)$ has a negative expectation value 
$\langle O_4 \rangle = - \frac{2 N_c + 3}{4N_c}$  (Table 
\ref{Matrix}) for flavour singlets,
 which  lowers the total mass,  but not enough. The fit considerably improves 
by the introduction of the operator $B_2$, considered in this work for the first time. 
As one can see, this operator helps in lowering the flavor singlets with respect to the rest of the spectrum. 

%Perhaps  the good agreement obtained in Ref. \cite{SGS} can also be explained by
%cancellation of various terms in the mass formula which contribute with alternating signs. 
%An important problem for future studies would be the physical understanding of the signs of coefficients 
%$c_i$ and $d_i$, besides their magnitudes. 

We have made some other fits which are not presented here. For example we have included the operators
$\frac{1}{N_c^2} L^{(2)}\cdot T \cdot\{ S, G\}$ and $\frac{3}{N^2_c} S \cdot T \cdot G$. 
As we found that their contributions to the global fit are negligible we have omitted them. 

Finally, our results are compatible with those of Cohen and Lebed \cite{COLEB2} where 
for SU(6) mixed symmetric spin-flavor multiplets five towers of 
states are predicted based on five independent operators: the $O_1$ operator of order $\mathcal{O}(N_c)$ 
and  four $\mathcal{O}(N^0_c)$ operators written in the core+excited quark scheme \cite{CCGL}.
In our case the latter operators correspond to  $O_2$, $O_4$, $O_5$ and $O_6$.  The  
operator $O_7$, also of order  $\mathcal{O}(N^0_c)$, has not been considered in \cite{COLEB2}, being 
more complex. 
All these are called quark-picture operators. It would be useful to reanalyze the connection 
 established by Cohen and Lebed  between the quark-picture operators and the K-matrix poles of their approach.
%This is a further step to tackle.

%%%%%%%%%%%%%%%%%%%%%%%%%%%%%%%%%%%%%%%%%%%%%%%%%%%%%%%%%%%%%%%%
\section{Conclusions}

We have analyzed the spectrum of nonstrange and strange baryons
belonging to the lowest negative parity 70-plet, to first order in SU(3) 
symmetry breaking by using a new method which in the $1/N_c$ expansion 
simplifies the mass formula, reducing it to a considerably smaller number of terms, namely 7 operators of type $O_i$
as compared to 11 operators  in Ref. \cite{CCGL}.
This allows us to  easier find the most dominant operators to order $1/N_c$.
We have shown that the isospin operator $O_4$, neglected in previous studies,
contributes to decuplets with a coefficient of the same order of magnitude as the spin 
operator $O_3$ in octets.  In addition the role of the operator $O_4$ is
crucial in describing the flavor singlets. 

Actually the remaining difficulty in perfectly fitting  $\Lambda (1405)$ is consistent with 
the view that this resonance has a
more complex nature  as, for example,  having a coupling to a $\bar K N$ system,
which might survive in the large $N_c$ limit \cite{GarciaRecio:2006wb,Hyodo:2007np}.
We remind that the meson-baryon coupling is a long standing problem discussed first in the Skyrme model
\cite{MattisMukerjee} and later, in the resonant picture of the meson-nucleon scattering \cite{COLEB1}.

The isoscalar factors found in this work can be used in  further SU(6) studies, formally or in 
physical applications.

%%%%%%%%%%%%%%%%%%%%%%%%%%%%%%%%%%%%%%%%%%%%%%%%%%%%%%%%%%%%%%%%%

\begin{sidewaystable}
\caption{Isoscalar factors of the SU(6) generators
 Eqs. (\ref{normes}) and (\ref{GEN}),
% $[N_c-1,1] \times [21^4] \rightarrow [N_c-1,1]$ 
corresponding to 
the $^28$ multiplet of $N_c = 3$.}
{\scriptsize
 \renewcommand{\arraystretch}{1.5}
\begin{tabular}{l|c|c|l}
\hline
\hline
$(\lambda_1\mu_1)S_1$ \hspace{0.5cm} & \hspace{0.5cm}$(\lambda_2\mu_2)S_2$ \hspace{0.5cm} & \hspace{0.5cm}$\rho$\hspace{0.5cm} & \hspace{0.5cm}$\left(\begin{array}{cc||c}                                         [N_c-1,1]  &  [21^4]  &  [N_c-1,1] \\
                           (\lambda_1\mu_1)S_1 & (\lambda_2\mu_2)S_2 & (\lambda\mu)S
                                      \end{array}\right)_\rho$  \\
\vspace{-0.6cm} &  &   & \\
\hline
$(\lambda\mu)S+1$\hspace{0.5cm} & \hspace{0.cm}$(11)1$ & $1$ &\hspace{0.5cm}$-\frac{3\sqrt{2S(2S+3)(N_c+2S+2)}}{\sqrt{(S+1)(2S+1)\left[N_c(N_c+6)+12S(S+1)\right](5N_c+18)}}$\\
$(\lambda\mu)S+1$\hspace{0.5cm} & \hspace{0.cm}$(11)1$ & $2$ &\hspace{0.5cm}$\frac{N_c}{S+1}\sqrt{\frac{3(2S+3)(N_c-2S+4)(N_c+2S+6)}{2(2S+1)(N_c-2S)\left[N_c(N_c+6)+12S(S+1)\right](5N_c+18)}}$\\
$(\lambda\mu)S$\hspace{0.5cm} & \hspace{0.cm}$(11)1$ & $1$ &\hspace{0.5cm}$\left\{12S(S+1)+N_c[4S(S+1)-3]\right\}\sqrt{\frac{2}{S(S+1)\left[N_c(N_c+6)+12S(S+1)\right]N_c(5N_c+18)}}$\\
$(\lambda\mu)S$\hspace{0.5cm} & \hspace{0.cm}$(11)1$ & $2$ &\hspace{0.5cm}$\frac{4S^2(S+1)^2-2N_cS(S+1)-(S^2+S-1)N_c^2}{2S(S+1)}\sqrt{\frac{6(N_c-2S+4)(N_c+2S+6)}{(N_c-2S)(N_c+2S+2)\left[N_c(N_c+6)+12S(S+1)\right]N_c(5N_c+18)}}$\\
$(\lambda\mu)S-1$\hspace{0.5cm} & \hspace{0.cm}$(11)1$ & $1$ &\hspace{0.5cm}$-3\sqrt{\frac{2(S+1)(2S-1)(N_c-2S)}{S(2S+1)(N_c(N_c+6)+12S(S+1))(5N_c+18)}}$ \\
$(\lambda\mu)S-1$\hspace{0.5cm} & \hspace{0.cm}$(11)1$ & $2$ &\hspace{0.5cm}$\frac{N_c}{S}\sqrt{\frac{3(2S-1)(N_c-2S+4)(N_c+2S+6)}{2(2S+1)(N_c+2S+2)(N_c(N_c+6)+12S(S+1))(5N_c+18)}}$ \\
$(\lambda+2,\mu-1)S+1$\hspace{0.5cm} & \hspace{0.cm}$(11)1$ & $/$ &\hspace{0.5cm}$-\frac{1}{S+1}\sqrt{\frac{3S(S+2)(2S+3)(N_c-2S-2)(N_c+2S+2)(N_c+2S+6)}{2(2S+1)(N_c+2S+4)N_c(5N_c+18)}}$\\
$(\lambda+2,\mu-1)S$\hspace{0.5cm} & \hspace{0.cm}$(11)1$ & $/$ &\hspace{0.5cm}$\frac{1}{S+1}\sqrt{\frac{3(2S+3)(N_c+2S+2)(N_c+2S+6)}{2(2S+1)(N_c+2S+4)(5N_c+18)}}$ \\
$(\lambda+1,\mu-2)S+1$\hspace{0.5cm} & \hspace{0.cm}$(11)1$ & $/$ &\hspace{0.5cm}$-2\sqrt{\frac{3S(2S+3)(N_c-2S-2)}{(S+1)(2S+1)(N_c-2S)(N_c+2S+4)(5N_c+18)}}$ \\
$(\lambda+1,\mu-2)S$\hspace{0.5cm} & \hspace{0.cm}$(11)1$ & $/$ &\hspace{0.5cm}$-2\sqrt{\frac{3(N_c-2S-2)}{(S+1)(2S+1)(N_c-2S)(N_c+2S+4)(5N_c+18)}}$ \\
$(\lambda-1,\mu-1)S$\hspace{0.5cm} & \hspace{0.cm}$(11)1$ & $/$ &\hspace{0.5cm}$\sqrt{\frac{12(N_c+2S)}{S(2S+1)(N_c-2S+2)(N_c+2S+2)(5N_c+18)}}$\\
$(\lambda-1,\mu-1)S-1$\hspace{0.5cm} & \hspace{0.cm}$(11)1$ & $/$ &\hspace{0.5cm}$-2\sqrt{\frac{3(S+1)(N_c+2S)(2S-1)}{S(2S+1)(N_c-2S+2)(N_c+2S+2)(5N_c+18)}}$\\
$(\lambda-2,\mu+1)S$\hspace{0.5cm} & \hspace{0.cm}$(11)1$ & $/$ &\hspace{0.5cm}$\frac{1}{S}\sqrt{\frac{3(2S-1)(N_c-2S)(N_c-2S+4)}{2(2S+1)(N_c-2S+2)(5N_c+18)}}$\\
$(\lambda-2,\mu+1)S-1$\hspace{0.5cm} & \hspace{0.cm}$(11)1$ & $/$ &\hspace{0.5cm}$-\frac{1}{S}\sqrt{\frac{3(S-1)(S+1)(2S-1)(N_c-2S)(N_c+2S)(N_c-2S+4)}{2(2S+1)(N_c-2S+2)N_c(5N_c+18)}}$ \\
$(\lambda\mu)S$\hspace{0.5cm} & \hspace{0.cm}$(11)0$ & $1$ &\hspace{0.5cm}$\sqrt{\frac{N_c(N_c+6)+12S(S+1)}{2N_c(5N_c+18)}}$\\
$(\lambda\mu)S$\hspace{0.5cm} & \hspace{0.cm}$(11)0$ & $2$ &\hspace{0.5cm} 0\\
$(\lambda\mu)S$\hspace{0.5cm} & \hspace{0.cm}$(00)1$ & $/$ &\hspace{0.5cm} $\sqrt{\frac{4S(S+1)}{N_c(5N_c+18)}}$\\ 
\hline
\hline
\end{tabular}}
%\caption{Isoscalar factors SU(6) "utiles" for $[N_c-1,1]
% \times [21^4] \rightarrow [N_c-1,1]$ final state $^28$.}
\label{octet_spin_one_half} 
\end{sidewaystable}
%%%%%%%%%%%%%%%%%%%%%%%%%%%%%%%%%%%%%
\begin{sidewaystable}
\caption{Isoscalar factors of the SU(6) generators,
% $[N_c-1,1] \times [21^4] \rightarrow [N_c-1,1]$ 
corresponding to 
the $^48$ multiplet of $N_c = 3$.}
{\scriptsize
 \renewcommand{\arraystretch}{1.5}
\begin{tabular}{l|c|c|l}
\hline
\hline
$(\lambda_1\mu_1)S_1$ \hspace{0.5cm} & \hspace{0.5cm}$(\lambda_2\mu_2)S_2$ \hspace{0.5cm} & \hspace{0.5cm}$\rho$\hspace{0.5cm} & \hspace{0.5cm}$\left(\begin{array}{cc||c}                                         [N_c-1,1]  &  [21^4]  &  [N_c-1,1] \\
                           (\lambda_1\mu_1)S_1 & (\lambda_2\mu_2)S_2 & (\lambda-2,\mu+1)S
                                      \end{array}\right)_\rho$  \\
\vspace{-0.6cm} &  &   & \\
\hline
$(\lambda-2,\mu+1)S$\hspace{0.5cm} & \hspace{0.cm}$(11)1$ & $1$ &\hspace{0.5cm}$ \left[N_c(4S-3)+6S\right]\sqrt{\frac{2(S+1)}{S\left[N_c(N_c+6)+12(S-1)S\right]N_c(5N_c+18)}}$\\
$(\lambda-2,\mu+1)S$\hspace{0.5cm} & \hspace{0.cm}$(11)1$ & $2$ &\hspace{0.5cm}$-\frac{N_c-2S}{S}\sqrt{\frac{3(S-1)(S+1)(N_c-2S+6)(N_c+2S)(N_c+2S+4)}{2(N_c-2S+2)\left[N_c(N_c+6)+12(S-1)S\right]N_c(5N_c+18)}}$\\
$(\lambda\mu)S+1$\hspace{0.5cm} & \hspace{0.cm}$(11)1$ & $/$ &\hspace{0.5cm}$-\sqrt{\frac{3}{2}}\sqrt{\frac{2S+3}{2S+1}}\sqrt{\frac{(N_c-2S)(N_c+2S+4)}{N_c(5N_c+18)}}$\\
$(\lambda\mu)S$\hspace{0.5cm} & \hspace{0.cm}$(11)1$ & $/$ &\hspace{0.5cm}$-\frac{1}{S}\sqrt{\frac{3}{2}}\sqrt{\frac{(N_c-2S)(N_c+2S+4)}{(N_c+2S+2)(5N_c+18)}}$\\
$(\lambda\mu)S-1$\hspace{0.5cm} & \hspace{0.cm}$(11)1$ & $/$ &\hspace{0.5cm}$\frac{N_c+4S^2}{S}\sqrt{\frac{3(N_c+2S+4)}{2(2S-1)(2S+1)(N_c+2S+2)N_c(5N_c+18)}}$\\
$(\lambda-2,\mu+1)S-1$\hspace{0.5cm} & \hspace{0.cm}$(11)1$ & $1$ &\hspace{0.5cm}$\frac{3\sqrt{2(S-1)(N_c+2S)}}{\sqrt{S\left[N_c(N_c+6)+12(S-1)S\right](5N_c+18)}}$ \\
$(\lambda-2,\mu+1)S-1$\hspace{0.5cm} & \hspace{0.cm}$(11)1$ & $2$ &\hspace{0.5cm}$-\frac{N_c}{S}\sqrt{\frac{3(N_c-2S+6)(N_c+2S+4)}{2(N_c-2S+2)\left[N_c(N_c+6)+12(S-1)S\right](5N_c+18)}}$\\
$(\lambda-1,\mu-1)S$\hspace{0.5cm} & \hspace{0.cm}$(11)1$ & $/$ &\hspace{0.5cm}$-2\sqrt{\frac{3(S+1)(N_c-2S)(N_c+2S)}{S(N_c-2S+2)(N_c+2S+2)N_c(5N_c+18)}}$\\
$(\lambda-1,\mu-1)S-1$\hspace{0.5cm} & \hspace{0.cm}$(11)1$ & $/$ &\hspace{0.5cm}$2(S-1)\sqrt{\frac{3(N_c-2S)(N_c+2S)}{S(2S-1)(N_c-2S+2)(N_c+2S+2)N_c(5N_c+18)}}$\\
$(\lambda-3,\mu)S-1$\hspace{0.5cm} & \hspace{0.cm}$(11)1$ & $/$ &\hspace{0.5cm}$-2\sqrt{\frac{3(S-1)(N_c+2S-2)}{(2S-1)(N_c-2S+4)N_c(5N_c+18)}}$\\
$(\lambda-4,\mu+2)S-1$\hspace{0.5cm} & \hspace{0.cm}$(11)1$ & $/$ &\hspace{0.5cm}$-\sqrt{\frac{3}{2}}\sqrt{\frac{2S-3}{2S-1}}\sqrt{\frac{(N_c+2S)(N_c-2S+2)(N_c-2S+6)}{(N_c-2S+4)N_c(5N_c+18)}}$\\
$(\lambda-2,\mu+1)S$\hspace{0.5cm} & \hspace{0.cm}$(11)0$ & $1$ &\hspace{0.5cm}$\sqrt{\frac{N_c(N_c+6)+12(S-1)S}{2N_c(5N_c+18)}}$\\
$(\lambda-2,\mu+1)S$\hspace{0.5cm} & \hspace{0.cm}$(11)0$ & $2$ &\hspace{0.5cm}$0$\\
$(\lambda-2,\mu+1)S$\hspace{0.5cm} & \hspace{0.cm}$(00)1$ & $/$ &\hspace{0.5cm}$\sqrt{\frac{4S(S+1)}{N_c(5N_c+18)}}$\\
\hline
\hline
\end{tabular}}
\label{octet_spin_three_halfs} 
\end{sidewaystable}

\begin{sidewaystable}
\caption{Isoscalar factors of the SU(6) generators,
% $[N_c-1,1] \times [21^4] \rightarrow [N_c-1,1]$ 
corresponding to 
the $^210$ multiplet of $N_c = 3$.}
{\scriptsize
 \renewcommand{\arraystretch}{1.5}
\begin{tabular}{l|c|c|l}
\hline
\hline
$(\lambda_1\mu_1)S_1$ \hspace{0.5cm} & \hspace{0.5cm}$(\lambda_2\mu_2)S_2$ \hspace{0.5cm} & \hspace{0.5cm}$\rho$\hspace{0.5cm} & \hspace{0.5cm}$\left(\begin{array}{cc||c}                                         [N_c-1,1]  &  [21^4]  &  [N_c-1,1] \\
                           (\lambda_1\mu_1)S_1 & (\lambda_2\mu_2)S_2 & (\lambda+2,\mu-1)S
                                      \end{array}\right)_\rho$  \\
\vspace{-0.6cm} &  &   & \\
\hline
$(\lambda+4,\mu-2)S+1$\hspace{0.5cm} & \hspace{0.cm}$(11)1$ & $/$ &\hspace{0.5cm}$-\sqrt{\frac{3(2S+5)(N_c+2S+4)(N_c+2S+8)(N_c-2S-2)}{2(2S+3)(N_c+2S+6)N_c(5N_c+18)}}$ \\
$(\lambda+3,\mu-3)S+1$\hspace{0.5cm} & \hspace{0.cm}$(11)1$ & $/$ &\hspace{0.5cm}$2\sqrt{\frac{3(S+2)(N_c-2S-4)}{(2S+3)(N_c+2S+6)N_c(5N_c+18)}}$\\
$(\lambda+1,\mu-2)S+1$\hspace{0.5cm} & \hspace{0.cm}$(11)1$ & $/$ &\hspace{0.5cm}$-2(S+2)\sqrt{\frac{3(N_c+2S+2)(N_c-2S-2)}{(S+1)(2S+3)(N_c-2S)(N_c+2S+4)N_c(5N_c+18)}}$\\
$(\lambda+2,\mu-1)S+1$\hspace{0.5cm} & \hspace{0.cm}$(11)1$ & $1$ &\hspace{0.5cm}$3\sqrt{\frac{2(S+2)(N_c-2S-2)}{(S+1)(N_c(N_c+6)+12(S+1)(S+2))(5N_c+18)}}$\\
$(\lambda+2,\mu-1)S+1$\hspace{0.5cm} & \hspace{0.cm}$(11)1$ & $2$ &\hspace{0.5cm}$-\frac{N_c}{S+1}\sqrt{\frac{3(N_c-2S+2)(N_c+2S+8)}{2(N_c+2S+4)(N_c(N_c+6)+12(S+1)(S+2))(5N_c+18)}}$\\
$(\lambda+2,\mu-1)S$\hspace{0.5cm} & \hspace{0.cm}$(11)1$ & $1$ &\hspace{0.5cm}$\left[N_c(4S+7)+6(S+1)\right]\sqrt{\frac{2S}{(S+1)\left[N_c(N_c+6)+12(S+1)(S+2)\right]N_c(5N_c+18)}}$\\
$(\lambda+2,\mu-1)S$\hspace{0.5cm} & \hspace{0.cm}$(11)1$ & $2$ &\hspace{0.5cm}$-\frac{N_c+2S+2}{S+1}\sqrt{\frac{3S(S+2)(N_c-2S-2)(N_c-2S+2)(N_c+2S+8)}{2(N_c+2S+4)\left[N_c(N_c+6)+12(S+1)(S+2)\right]N_c(5N_c+18)}}$\\
$(\lambda+1,\mu-2)S$\hspace{0.5cm} & \hspace{0.cm}$(11)1$ & $/$ &\hspace{0.5cm}$2\sqrt{\frac{3S(N_c+2S+2)(N_c-2S-2)}{(S+1)(N_c-2S)(N_c+2S+4)N_c(5N_c+18)}}$\\
$(\lambda\mu)S+1$\hspace{0.5cm} & \hspace{0.cm}$(11)1$ & $/$ &\hspace{0.5cm}$\frac{N_c+4(S+1)^2}{S+1}\sqrt{\frac{3(N_c-2S+2)}{2(2S+1)(2S+3)(N_c-2S)N_c(5N_c+18)}}$ \\
$(\lambda\mu)S$\hspace{0.5cm} & \hspace{0.cm}$(11)1$ & $/$ &\hspace{0.5cm} $-\frac{1}{S+1}\sqrt{\frac{3(N_c+2S+2)(N_c-2S+2)}{2(N_c-2S)(5N_c+18)}}$\\
$(\lambda\mu)S-1$\hspace{0.5cm} & \hspace{0.cm}$(11)1$ & $/$ &\hspace{0.5cm} $-\sqrt{\frac{3(2S-1)(N_c+2S+2)(N_c-2S+2)}{2(2S+1)N_c(5N_c+18)}}$\\
$(\lambda+2,\mu-1)S$\hspace{0.5cm} & \hspace{0.cm}$(11)0$ & $1$ &\hspace{0.5cm}$\sqrt{\frac{N_c(N_c+6)+12(S+1)(S+2)}{2N_c(5N_c+18)}}$\\
$(\lambda+2,\mu-1)S$\hspace{0.5cm} & \hspace{0.cm}$(11)0$ & $2$ &\hspace{0.5cm} 0\\
$(\lambda+2,\mu-1)S$\hspace{0.5cm} & \hspace{0.cm}$(00)1$ & $/$ &\hspace{0.5cm} $\sqrt{\frac{4S(S+1)}{N_c(5N_c+18)}}$\\
\hline
\hline
\end{tabular}}
%\caption{Isoscalar factors SU(6) "utiles" for 
%$[N_c-1,1] \times [21^4] \rightarrow [N_c-1,1]$ final state $^210$.}

\label{decuplet_spin_one_half} 
\end{sidewaystable}

%%%%%%%%%%%%%%%%%%%%%%%%%%%%%%%%%%%%%%%%%%%%%%%%%%%%%%%%%%%%%%%%%%%%%%%%%%
\begin{sidewaystable}
\caption{Isoscalar factors of the SU(6) generators,
% $[N_c-1,1] \times [21^4] \rightarrow [N_c-1,1]$ 
corresponding to 
the $^21$ multiplet of $N_c = 3$.}
{\scriptsize
 \renewcommand{\arraystretch}{1.5}
\begin{tabular}{l|c|c|l}
\hline
\hline
$(\lambda_1\mu_1)S_1$ \hspace{0.5cm} & \hspace{0.5cm}$(\lambda_2\mu_2)S_2$ \hspace{0.5cm} & \hspace{0.5cm}$\rho$\hspace{0.5cm} & \hspace{0.5cm}$\left(\begin{array}{cc||c}                                         [N_c-1,1]  &  [21^4]  &  [N_c-1,1] \\
                           (\lambda_1\mu_1)S_1 & (\lambda_2\mu_2)S_2 & (\lambda-1,\mu-1)S
                                      \end{array}\right)_\rho$  \\
\vspace{-0.6cm} &  &   & \\
\hline
$(\lambda+1,\mu-2)S+1$\hspace{0.5cm} & \hspace{0.cm}$(11)1$ & $/$ &\hspace{0.5cm}$-\sqrt{\frac{3(2S+3)(N_c-2S-2)(N_c-2S+2)(N_c+2S+4)}{2(N_c-2S)(2S+1)N_c(5N_c+18)}}$\\
$(\lambda+1,\mu-2)S$\hspace{0.5cm} & \hspace{0.cm}$(11)1$ & $/$ &\hspace{0.5cm}$-\sqrt{\frac{3(N_c-2S-2)(N_c-2S+2)(N_c+2S+4)}{2S(2S+1)(N_c-2S)N_c(5N_c+18)}}$\\
$(\lambda-1,\mu-1)S$\hspace{0.5cm} & \hspace{0.cm}$(11)1$ & $1$ &\hspace{0.5cm}$\left[N_c(4S-3)+6S\right]\sqrt{\frac{2(S+1)}{S\left[N_c^2+12(S^2-1)\right]N_c(5N_c+18)}}$\\
$(\lambda-1,\mu-1)S$\hspace{0.5cm} & \hspace{0.cm}$(11)1$ & $2$ &\hspace{0.5cm}$-\left\{N_c(N_c+6)-4\left[S(S-1)-3\right]\right\}\sqrt{\frac{3(2S-1)(S+1)(N_c-2S-2)(N_c+2S-2)}{2S(2S+1)(N_c-2S+2)(N_c+2S+2)\left[N_c^2+12(S^2-1)\right]N_c(5N_c+18)}}$\\
$(\lambda-1,\mu-1)S-1$\hspace{0.5cm} & \hspace{0.cm}$(11)1$ & $1$ &\hspace{0.5cm}$3\sqrt{\frac{2N_c(2S-1)}{S[N_c^2+12(S^2-1)](5N_c+18)}}$\\
$(\lambda-1,\mu-1)S-1$\hspace{0.5cm} & \hspace{0.cm}$(11)1$ & $2$ &\hspace{0.5cm} 0 \hspace{0.5cm} if $S=1/2$\\
$(\lambda-1,\mu-1)S-1$\hspace{0.5cm} & \hspace{0.cm}$(11)1$ & $2$ &\hspace{0.5cm} $-\left[N_c(N_c+6)-12(S^2-1)\right]\sqrt{\frac{3(N_c-2S-2)(N_c+2S-2)}{2S(2S+1)(N_c-2S+2)(N_c+2S+2)[N_c^2+12(S^2-1)]N_c(5N_c+18)}}$ \hspace{0.5cm} if $S\geq 1$ \\
$(\lambda\mu)S+1$\hspace{0.5cm} & \hspace{0.cm}$(11)1$ & $/$ &\hspace{0.5cm}$\sqrt{\frac{6(2S+3)(N_c+2S+4)}{(2S+1)(N_c-2S)N_c(5N_c+18)}}$\\
$(\lambda\mu)S$\hspace{0.5cm} & \hspace{0.cm}$(11)1$ & $/$ &\hspace{0.5cm}$\frac{1}{S}\sqrt{\frac{6(N_c+2S+4)}{(N_c-2S)(N_c+2S+2)(5N_c+18)}}$\\
$(\lambda\mu)S-1$\hspace{0.5cm} & \hspace{0.cm}$(11)1$ & $/$ &\hspace{0.5cm}$\frac{S+1}{S}\sqrt{\frac{6(2S-1)(N_c+2S+4)}{(2S+1)(N_c+2S+2)N_c(5N_c+18)}}$\\
$(\lambda-2,\mu+1)S$\hspace{0.5cm} & \hspace{0.cm}$(11)1$ & $/$ &\hspace{0.5cm} $\frac{1}{S}\sqrt{\frac{6(S+1)(N_c-2S+4)(2S-1)}{(N_c-2S+2)N_c(5N_c+18)}}$\\
$(\lambda-2,\mu+1)S-1$\hspace{0.5cm} & \hspace{0.cm}$(11)1$ & $/$ &\hspace{0.5cm} $\frac{1}{S}\sqrt{\frac{6(N_c-2S+4)(S-1)(2S-1)}{(N_c-2S+2)(N_c+2S)(5N_c+18)}}$\\
$(\lambda-3,\mu)S-1$\hspace{0.5cm} & \hspace{0.cm}$(11)1$ & $/$ &\hspace{0.5cm} 0\hspace{0.5cm}  if $S=1/2$\\
$(\lambda-3,\mu)S-1$\hspace{0.5cm} & \hspace{0.cm}$(11)1$ & $/$ &\hspace{0.5cm} $-\sqrt{\frac{3(N_c+2S-2)(N_c+2S+2)(N_c-2S+4)(S-1)}{2S(N_c+2S)N_c(5N_c+18)}}$ \hspace{0.5cm} if $S\geq 1$ \\
$(\lambda-1,\mu-1)S$\hspace{0.5cm} & \hspace{0.cm}$(11)0$ & $1$ &\hspace{0.5cm}$\sqrt{\frac{N_c^2+12(S^2-1)}{2N_c(5N_c+18)}}$\\
$(\lambda-1,\mu-1)S$\hspace{0.5cm} & \hspace{0.cm}$(11)0$ & $2$ &\hspace{0.5cm} 0 \\
$(\lambda-1,\mu-1)S$\hspace{0.5cm} & \hspace{0.cm}$(00)1$ & $/$ &\hspace{0.5cm} $\sqrt{\frac{4S(S+1)}{N_c(5N_c+18)}}$\\ 
\hline
\hline
\end{tabular}}
%\caption{Isoscalar factors SU(6) "utiles" for 
%$[N_c-1,1] \times [21^4] \rightarrow [N_c-1,1]$ final state $^21$.}
\label{singlet_spin_one_half} 
\end{sidewaystable}

%%%%%%%%%%%%%%%%%%%%%%%%%%%%%%%%%%%%%%%%%%%%%%%%%%%%%%%%%%%%%%%
\appendix

\section{}
We recall that the  isoscalar factors of SU(3) obey the following 
orthogonality relation
\begin{widetext}
\begin{equation}
\sum_{Y'' I'' Y^a I^a}
 \left(\begin{array}{cc||c}
	(\lambda'' \mu'')    &  (11)   & (\lambda' \mu')   \\
	 Y'' I'' &  Y^a I^a   &   Y I
      \end{array}\right)_{\rho}
   \left(\begin{array}{cc||c}
	 (\lambda'' \mu'')   &  (11)   & (\lambda \mu) \\
	Y'' I''  &   Y^a I^a  &   Y' I'
      \end{array}\right)_{\rho} = \delta_{\lambda' \lambda} 
      \delta_{\mu' \mu} \delta_{Y' Y}\delta_{I' I},   
 \end{equation}
 \end{widetext}
 which can be easily checked. 
For completeness also note that the isoscalar factors obey the 
following symmetry property

%{\tiny
%\begin{tabular}{c}
%\vspace{-0.2cm}  \\
%\begin{eqnarray}
%\begin{widetext}
\begin{eqnarray}
\lefteqn{ \left(\begin{array}{cc||c}  (\lambda \mu)  &  (11)  &  (\lambda' \mu') \\
                              YI            &  -Y^aI^a & Y'I'
                                      \end{array}\right)=} \nonumber \\  & & 
(-)^{\frac{1}{3}(\mu'-\mu-\lambda'+\lambda+\frac{3}{2}Y^a)+I'-I}
\sqrt{\frac{\mathrm{dim}(\lambda'\mu')(2I+1)}{\mathrm{dim}(\lambda\mu)(2I'+1)}}
\left(\begin{array}{cc||c}   

                             (\lambda' \mu')  &  (11)  &  (\lambda \mu) \\
                              Y'I' & Y^aI^a &  YI 
                                      \end{array}\right).
\end{eqnarray}
%\end{widetext}				      
where $\mathrm{dim}(\lambda\mu) = \frac{1}{2}(\lambda+1)(\mu+1)(\lambda+\mu+2)$ 
is the dimension of the irrep $(\lambda\mu)$ of SU(3).			      
%\end{tabular}}
%\end{eqnarray}
The SU(6) isoscalar factors satisfy to the following symmetry property:
\begin{eqnarray}
 \lefteqn{\left(\begin{array}{cc||c}   [f] &  [21^4]  & [f]  \\
                                             (\lambda_1 \mu_1)S_1 & (\lambda_2\mu_2)S_2 &  (\lambda\mu)S
                                      \end{array}\right) =} \nonumber \\ & & 
              (-1)^{1/3(\mu_1 -\mu -\lambda_1 +\lambda)}(-1)^{S_1-S}\sqrt{\frac{\mathrm{dim}(\lambda_1\mu_1)(2S_1+1)}{\mathrm{dim}(\lambda\mu)(2S+1)}}
 \left(\begin{array}{cc||c}   [f] &  [21^4]  & [f]  \\
                                             (\lambda \mu)S & (\lambda_2\mu_2)S_2 &  (\lambda_1\mu_1)S_1
                                      \end{array}\right).              
\end{eqnarray}

%%%%%%%%%%%%%%%%%%%%%%%%%%%%%%%%%%%%%%%%%%%%%%%%%%%%%%%%%%%

%%%%%%%%%%%%%%%%%%%%%%%%%%%%%%%%%%%%%%%%%%%%%%%%%%%%%%%%%%%%%%%%
\section{}
Here we present the analytic form of the matrix elements of operators proportional to
$O_5$, $O_6$. They have been obtained following the approach described in Sec. III.
We recall that we have the following notations
$C^{SU(3)}_{(\lambda\mu)} = \frac{1}{3}(\lambda^2+\mu^2 +\lambda\mu
+3\lambda+3\mu)$ is the Casimir operator of the SU(3) irrep $(\lambda\mu)$
and  $C^{SU(6)}_{[f]}$ is the Casimir operator of the SU(6) irrep $[f]$.
In the present case we have $[f] = [N_c-1,1]$ so that
\begin{equation}
C^{SU(6)}_{[f]} = \frac{N_c(5 N_c + 18)}{12}.
\end{equation}
%%%%%%%%%%%%%%%%%%%%%%%%%%%%%%%%%%%%%%%%%%%%%%%%%%%%%%%%%%

\noindent
Up to the factor $\frac{3}{N_c}$ the matrix elements of the operator $O_5$ are obtained
from the matrix elements of $L \cdot T \cdot G$ for which we have obtained the 
following form 
\begin{widetext}
\begin{eqnarray}
\lefteqn{ \langle (\lambda' \mu') Y'I'I_3'; \ell'S'JJ_3
 |\sum_i (-1)^{i+a}L^iT^aG^{-(ia)}|(\lambda \mu) YII_3; \ell SJJ_3\rangle =}\nonumber \\ & &   \delta_{\ell'\ell} \delta_{\lambda\lambda'}\delta_{\mu\mu'}\delta_{Y'Y}\delta_{I'I} \delta_{I_3'I_3} 
   (-1)^{J+\ell+S'}   C^{SU(6)}_{[f]}
 \sqrt{\ell(\ell+1)(2\ell+1)(2S'+1)}\nonumber \\ & &  \times\left\{
  \begin{array}{ccc}
   \ell & \ell & 1 \\
   S' & S & J
  \end{array}\right\}
%  \left(\begin{array}{cc||c}
%         [N_c-1,1] & [21^2] & [N_c-1,1] \\
%	 SI & 11 & S'I 
%        \end{array}\right)_1
  \left(\begin{array}{cc||c}
         [f]           & [21^4] & [f] \\
	 (\lambda\mu)S & (11)1 & (\lambda\mu)S' 
        \end{array}\right)_{\rho=1}
     \left(\begin{array}{cc||c}
         [f]           & [21^4] & [f] \\
	 (\lambda\mu)S & (11)0 & (\lambda\mu)S 
        \end{array}\right)_{\rho=1}.
 \end{eqnarray}
\end{widetext}
\noindent
where we have used the short-hand notation
\begin{equation}
(-)^a = (-)^{I^a_3 + \frac{Y^a}{2}}.
\end{equation}
In this way one can easily define the values of $\lambda$ and $\mu$ to be used in the
tables for the octets,  the decuplets and the singlets.  The reason is that when calculating 
the isoscalar factors we took $\lambda = 2 S$ and $\mu = \frac{N_c-2 S}{2}$. These are 
definitions consistent with the inner products in the flavor-spin space and provided the advantage
of expressing  the isoscalar factors in terms of the spin $S$ of a given state and $N_c$ only.
Let us take the example of the two octets $^28$ and $^48$. They represent the same flavor state
but of different spin. Then in using Table 1 of Ref.  \cite{Matagne:2008kb} one has to take 
$\lambda$ = 1 and $\mu = \frac{N_c-1}{2}$.  On the other hand in Table 2 or its extended form
Table \ref{octet_spin_three_halfs} from the next appendix, the flavor octet  $^48$ will be described
by the irrep $(\lambda-2,\mu+1)$ with $\lambda = 2 S = 3$ and  $\mu = \frac{N_c-3}{2}$
which gives the  irrep $(11)$ when $N_c = 3$, as it should be.

For the expectation  value of   $L^{2} \cdot G \cdot G$  we have obtained the following expression
%\begin{widetext}
 \begin{eqnarray}
 \lefteqn{  \langle (\lambda'\mu')Y'I'I_3';\ell'S' JJ_3
 |(-1)^{i+j+a}L^{(2)ij}G^{-ia}G^{-j,-a}|(\lambda\mu)YII_3;\ell SJJ_3\rangle =  }\\
% \delta_{J'J} \delta_{J_3J_3'} 
 & &   \delta_{\ell'\ell}\delta_{\lambda\lambda'}\delta_{\mu\mu'}
 \delta_{Y'Y} \delta_{I'I} \delta_{I_3'I_3} (-1)^{J+\ell -S}
  \frac{1}{2}C^{SU(6)}_{[f]} \nonumber \\
%  \sqrt{2S'+1} 
  \times & & \sqrt{\frac{5\ell(\ell+1)(2\ell-1)(2\ell+1)(2\ell+3)}{6}}
  \left\{\begin{array}{ccc}
   \ell & \ell & 2 \\
      S & S' & J
  \end{array}\right\}\nonumber \\
  \times 
  & & \sqrt{(2S+1)(2S'+1)} \sum_{S''}(-1)^{(S-S'')}\left\{\begin{array}{ccc}
   1 & 1 & 2 \\
   S & S' & S''
  \end{array}\right\} 
  \nonumber \\ \times & & \sum_{\rho,\lambda'',\mu''}
   \left(\begin{array}{cc||c}
         [f] & [21^4] & [f] \\
	 (\lambda''\mu'')S'' & (11)1 & (\lambda\mu)S 
        \end{array}\right)_{\rho} 
    \left(\begin{array}{cc||c}
         [f] & [21^4] & [f] \\
	 (\lambda''\mu'')S'' & (11)1 & (\lambda\mu)S' 
        \end{array}\right)_{\rho}.
 \end{eqnarray} 
%\end{widetext}
which multiplied by the factor $\frac{15}{N_c}$  gives the expectation value of $O_6$.

Concerning $O_7$, one has
\begin{eqnarray}
 \lefteqn{  \langle (\lambda'\mu')Y'I'I_3';\ell'S' JJ_3
 |(-1)^{i+j+a}L^{i}G^{ja}G^{i,-a}S^{-j}|(\lambda\mu)YII_3;\ell SJJ_3\rangle =  }\nonumber \\
& &  (-1)^{J+\ell -S'}\frac{C^{SU(6)}_{[f]}}{2}(2S+1)\sqrt{\ell(\ell+ 1)(2\ell +1)}\sqrt{S(S+1)(2S'+1)}\nonumber \\
& & \times 
\left\{\begin{array}{ccc}
   1 & \ell & \ell \\
   J & S & S'
  \end{array}\right\}\ 
  \sum_{S''} 
  \left\{\begin{array}{ccc}
   S' & 1 & S \\
   S & 1 & S''
  \end{array}\right\} \nonumber \\
 & & \times \sum_{\lambda'',\mu'',\rho}
     \left(\begin{array}{cc||c}
         [f] & [21^4] & [f] \\
	 (\lambda''\mu'')S'' & (11)1 & (\lambda\mu)S 
        \end{array}\right)_{\rho} 
    \left(\begin{array}{cc||c}
         [f] & [21^4] & [f] \\
	 (\lambda''\mu'')S'' & (11)1 & (\lambda\mu)S' 
        \end{array}\right)_{\rho},
\end{eqnarray}
and
\begin{eqnarray}
\lefteqn{  \langle (\lambda'\mu')Y'I'I_3';\ell'S' JJ_3
 |(-1)^{i+j+a}L^{i}G^{ja}S^{-j}G^{-(ia)}|(\lambda\mu)YII_3;\ell SJJ_3\rangle =  }\nonumber \\
 & & (-1)^{\ell +S+J} \frac{C^{SU(6)}_{[f]}}{2} \sqrt{\ell(\ell +1)(2\ell +1)}\sqrt{S'(S'+1)(2S+1)} \nonumber\\
 & & \times 
\left\{\begin{array}{ccc}
   1 & \ell & \ell \\
   J & S & S'
  \end{array}\right\}\ 
  \sum_{\lambda'',\mu'',\rho}
     \left(\begin{array}{cc||c}
         [f] & [21^4] & [f] \\
	 (\lambda''\mu'')S' & (11)1 & (\lambda\mu)S 
        \end{array}\right)_{\rho} 
    \left(\begin{array}{cc||c}
         [f] & [21^4] & [f] \\
	 (\lambda''\mu'')S' & (11)1 & (\lambda\mu)S' 
        \end{array}\right)_{\rho}.
\end{eqnarray}

For completeness, we also give the general analytic form of the matrix elements
of $S\cdot T \cdot G$. This  is given by
\begin{eqnarray}
 \lefteqn{\langle 
 (\lambda'\mu')Y'I'I_3';\ell'S' JJ_3|(-1)^{i+a}S^iT^aG^{-(ia)}|(\lambda\mu)YII_3;\ell S JJ_3\rangle 
 =} \nonumber \\ 
 & &  \delta_{\ell'\ell} \delta_{S'S}
 \delta_{\lambda'\lambda} \delta_{\mu'\mu} \delta_{Y'Y} \delta_{I'I} 
 \delta_{I_3'I_3} 
 \sqrt{S(S+1)}
C^{SU(6)}_{[f]} \nonumber  \\
& & \times \sum_\rho \left(\begin{array}{cc||c}
         [f]           & [21^4] & [f] \\
	 (\lambda\mu)S & (11)0  & (\lambda\mu)S
        \end{array}\right)_{\rho}
  \left(\begin{array}{cc||c}
         [f]           & [21^4] & [f] \\
	 (\lambda\mu)S & (11)1  & (\lambda'\mu')S'
        \end{array}\right)_{\rho}.
\end{eqnarray}
The matrix elements of this operator are presented in Table \ref{Matrixstg}. 
 This operator is not considered in the analysis because it does not improve the fit.
\begin{table}
\begin{center}
\caption{Matrix elements of $S^iT^aG^{ia}$ for all states belonging to the 
$[{\bf 70},1^-]$ multiplet. the vanishing off-diagonal matrix elements are not indicated}
\label{Matrixstg}
\renewcommand{\arraystretch}{2.3} {\scriptsize
\begin{tabular}{lcc}
\hline
\hline
  & $S^iT^aG^{ia}$ \\
\hline
$^28_J$ & $\frac{3}{4}$ & \\
$^48_J$ & $\frac{5(N_c+3)}{8}$ &  \\
$^210_J$ &  $\frac{3(N_c+1)}{8}$  & \\
$^21_J$ & $0$ &  \\
\hline
\end{tabular}}
\end{center}
\end{table}

\section{}

Here we have completed the calculation of SU(6) isoscalar factors corresponding to the 
performed in Ref.  \cite{Matagne:2008kb}. Table \ref{octet_spin_one_half} contains eight more cases
of  $(\lambda_1,\mu_1) S_1$  corresponding to  $(\lambda,\mu) S-1$  with $\rho$ = 1, 2, 
$(\lambda+2,\mu-1) S+1$,  $(\lambda+1,\mu-2) S+1$,  $(\lambda+1,\mu-2) S$, 
$(\lambda-1,\mu-1) S-1$,  $(\lambda-2,\mu+1) S$ and $(\lambda-2,\mu+1) S-1$ respectively.
Table    \ref{octet_spin_three_halfs}  contains five more
cases corresponding   to $(\lambda_1,\mu_1) S_1$ =
$(\lambda,\mu) S+1$, $(\lambda,\mu) S$, $(\lambda-1,\mu-1) S$, $(\lambda-1,\mu-1) S-1$
and $(\lambda-4,\mu+2) S-1$. 
The latter case is not applicable to our physical problem. It has been derived for
testing the  orthonormalization  properties. In Table \ref{decuplet_spin_one_half} 
the rows 1-5, 8 and 11 are new.
In Table \ref{singlet_spin_one_half} the rows 1, 2,  5 - 7 and 10 - 14 are new.

%%%%%%%%%%%%%%%%%%%%%%%%%%%%%%%%%%%%%%%%%%%%%%%%%%%%%%%%%%%%%%%
\section{}

There are two  SU(3) breaking operators which are neglected in the fit because their contributions
are negligible. 
They are defined to have zero expectation values for non strange baryons, as follows 
\begin{equation}
\label{B3}
B_3 = \frac{1}{N_c} S^i G^{i8} - \frac{1}{2 \sqrt{3}} O_3
\end{equation}
and 
\begin{equation}
\label{B4}
B_4 = \frac{1}{N_c} L^i G^{i8} - \sqrt{\frac{3}{8}} O_2
\end{equation}

The general form of their matrix elements is obtained from the matrix elements of the following operators
\begin{eqnarray}
 \lefteqn{\langle 
 (\lambda'\mu')Y'I'I_3';\ell'S' JJ_3|(-1)^{i}S^iG^{-i8}|(\lambda\mu)YII_3;\ell S JJ_3\rangle 
 =   \delta_{\ell'\ell} \delta_{S'S}}\nonumber \\
  & &\times \sqrt{\frac{C^{SU(6)}_{[f]}}{2}}\sqrt{S(S+1)}
 \sum_{\rho}\left(\begin{array}{cc||c}
	 (\lambda\mu) & (11)  & (\lambda'\mu')\\
	 YI           &  00   &  YI
        \end{array}\right)_{\rho}
  \left(\begin{array}{cc||c}
         [f]           & [21^4] & [f] \\
	 (\lambda\mu)S & (11)1  & (\lambda'\mu')S
        \end{array}\right)_{\rho}.
\end{eqnarray}
and
\begin{eqnarray}
 \lefteqn{\langle 
 (\lambda'\mu')Y'I'I_3';\ell'S' JJ_3|(-1)^{i}L^iG^{-i8}|(\lambda\mu)YII_3;\ell S JJ_3\rangle 
 =}\nonumber \\ & &  \delta_{\ell'\ell} 
 (-1)^{\ell+S'+J}\sqrt{\frac{C^{SU(6)}_{[f]}}{2}}\sqrt{(2L+1)(2S'+1)}
 \left\{\begin{array}{ccc}
   S' & \ell & J \\
   \ell & S & 1
  \end{array}\right\} \nonumber \\
  & & \times \sum_{\rho}\left(\begin{array}{cc||c}
	 (\lambda\mu) & (11)  & (\lambda'\mu')\\
	 YI           &  00   &  YI
        \end{array}\right)_{\rho}
  \left(\begin{array}{cc||c}
         [f]           & [21^4] & [f] \\
	 (\lambda\mu)S & (11)1  & (\lambda'\mu')S'
        \end{array}\right)_{\rho}.
\end{eqnarray}

For general interest the explicit form of these matrix elements as a function of $N_c$ are given in Tables \ref{Matrixsg} and \ref{Matrixlg} respectively.
Here and elsewhere the notations $^{2S+1}8_J$, etc. represents the multiplicity ${2J+1}$ of the spin, the SU(3) multiplet 
in dimensional notation and $J$ is the total spin. Note that the matrix elements depend on $N_c$, the strangeness $\mathcal{S}$
and the isospin $I$. 

\begin{table}
\begin{center}
\caption{Matrix elements of $S^iG^{i8}$ for all states belonging to the 
$[{\bf 70},1^-]$ multiplet.}
\label{Matrixsg}
\renewcommand{\arraystretch}{2.3} {\scriptsize
\begin{tabular}{lcc}
\hline
\hline
  & $S^iG^{i8}$ \\
\hline
$^28_J$ & $\frac{9 - 36 I (I+1) + \mathcal{S}^2 (N_c-3)^2 - 6 N_c + N_c[-4 I (I+1) (N_c-6) + 9 N_c] - 
 2 \mathcal{S} [9 + N_c(N_c-18)]}{16 \sqrt{3} (N_c-1) (N_c+3)}$ & \\
$^48_J$ & $\frac{5[-\mathcal{S}^2(N_c-3)+4I(I+1)(N_c-3)+3(N_c+1)+2\mathcal{S}(N_c+3)]}{16\sqrt{3}(N_c-1)}$ & \\
$^210_J$ & $\frac{\sqrt{3}(1+\mathcal{S})}{8}$ & $N_c = 3$ \\
$^210_J$ & $\frac{2\mathcal{S} (N_c-3) - \mathcal{S}^2 (N_c+3) + 4 I (I+1) (N_c+3) - 
 3 (3N_c+5)}{16 \sqrt{3]} (N_c+5)}$ & $N_c > 3$ \\
$^21_J$ & $-\frac{\sqrt{3} (N_c-3)}{4(N_c+3)}$ &  \\
$^28_{J}-$ $^48_{J}$ & 0 &  \\
$^28_{J}-$ $^210_{J}$ & $-\frac{1}{16 \sqrt{3}}\sqrt{\frac{(5 - \mathcal{S} + 2 I) (1 + \mathcal{S} - 2 I) (-3 + \mathcal{S} + 2 I) (3 + \mathcal{S} + 2 I) N_c (N_c+3)}{( N_c-1) (N_c+5)}}$& \\
$^48_{J}-$ $^210_{J}$ & 0& \\
$^28_{J}-$ $^21_{J}$ & $\frac{3\sqrt{N_c}}{2(N_c+3)} $ & \\
$^48_{J}-$ $^21_{J}$ & 0 & \\
$^210_{J}-$ $^21_{J}$ & 0 & \\
\hline
\end{tabular}}
\end{center}
\end{table}

\begin{table}
\begin{center}
\caption{Matrix elements of $L^iG^{i8}$ for all states belonging to the 
$[{\bf 70},1^-]$ multiplet.}
\label{Matrixlg}
\renewcommand{\arraystretch}{1.5} {\scriptsize
\begin{tabular}{lcc}
\hline
\hline
  & $L^iG^{i8}$ & \\
\hline
$^28_{1/2}$ & $-\frac{72 (N_c-1) (N_c+3\mathcal{S}) + (N_c-3)^2 [3 - 2 \mathcal{S} (N_c-9) + 9 N_c + 
     \mathcal{S}^2 (N_c+3) - 4 I (1 + I) (N_c+3)]}{
 12 \sqrt{6} (N_c-1) (N_c+3)^2}$ & \\
$^48_{1/2}$ & $-\frac{5 (-\mathcal{S}^2 (N_c-3) + 4 I (1 + I) (N_c-3) + 3 (N_c+1) + 
    2 \mathcal{S} (N_c+3)}{24 \sqrt{6} (N_c-1)}$ &  \\
$^28_{3/2}$ & $\frac{72 (N_c-1) (N_c+3\mathcal{S}) + (N_c-3)^2 [3 - 2 \mathcal{S} (N_c-9) + 9 N_c + 
    \mathcal{S}^2 (N_c+3) - 4 I (1 + I) (N_c+3)]}{24 \sqrt{6} (N_c-1) (N_c+3)^2}$ & \\
$^48_{3/2}$ & $-\frac{-\mathcal{S}^2 (N_c-3) + 4 I (1 + I) (N_c-3) + 3 (N_c+1) + 2 \mathcal{S} (N_c+3)}{
 12 \sqrt{6} (N_c-1)}$ & \\
$^48_{5/2}$ & $\frac{-\mathcal{S}^2 (N_c-3) + 4 I (1 + I) (N_c-3) + 3 (N_c+1) + 
 2 \mathcal{S} (N_c+3)}{8 \sqrt{6} (N_c-1)}$ & \\
$^210_{1/2}$ & $\frac{15 - 2 \mathcal{S} (N_c-3) + 9 N_c + \mathcal{S}^2 (N_c+3) - 
 4 I (1 + I) (N_c+3)}{12 \sqrt{6} (N_c+5)}$ & \\
$^210_{3/2}$ & $\frac{2 \mathcal{S} ( N_c -3) - \mathcal{S}^2 (N_c +3) + 4 I (1 + I) (N_c+3) - 
 3 (3 N_c+5)}{24 \sqrt{6} (N_c +5)}$ & \\ 
$^21_{1/2}$ & $\frac{N_c-3}{\sqrt{6} (N_c + 3)}$ & \\
$^21_{3/2}$ & $-\frac{N_c-3}{2 \sqrt{6}(N_c+3)}$ & \\
$^28_{1/2}-$ $^48_{1/2}$ & $\frac{6 \mathcal{S} + [3 + (\mathcal{S}-2) \mathcal{S} - 4 I (1 + I)] N_c}{6(N_c-1)}\sqrt{\frac{N_c}{6(N_c+3)}}$ & \\
$^28_{3/2}-$ $^48_{3/2}$ & $\frac{6 \mathcal{S} + [3 + (\mathcal{S}-2) \mathcal{S} - 4 I (1 + I)] N_c}{12(N_c-1)}\sqrt{\frac{5N_c}{3(N_c+3)}}$ & \\
$^28_{1/2}-$ $^210_{1/2}$ & $\frac{1}{12}\sqrt{\frac{((2I+5-\mathcal{S}) (1 + \mathcal{S} - 2 I) (-3 + \mathcal{S} + 2 I) (3 + \mathcal{S} +  2 I) N_c (N_c+3)}{6( N_c-1) (N_c+5)}}$ & \\
$^28_{3/2}-$ $^210_{3/2}$ & $-\frac{1}{24}\sqrt{\frac{((2I+5-\mathcal{S}) (1 + \mathcal{S} - 2 I) (-3 + \mathcal{S} + 2 I) (3 + \mathcal{S} +  2 I) N_c (N_c+3)}{6( N_c-1) (N_c+5)}}$ & \\
$^48_{1/2}-$ $^210_{1/2}$ & $-\frac{N_c+9}{96}\sqrt{\frac{(2I+5 - \mathcal{S}) (1 + \mathcal{S} - 2 I) (-3 + \mathcal{S} + 2 I) (3 + \mathcal{S} + 2 I)}{3(N_c-1) (N_c+5)}}$ & \\
$^48_{3/2}-$ $^210_{3/2}$ & $-\frac{N_c+9}{96}\sqrt{\frac{5(2I+5 - \mathcal{S}) (1 + \mathcal{S} - 2 I) (-3 + \mathcal{S} + 2 I) (3 + \mathcal{S} + 2 I)}{6(N_c-1) (N_c+5)}}$ & \\
$^28_{1/2}-$ $^21_{1/2}$ & $-\frac{\sqrt{2N_c}}{N_c+3}$ & \\
$^28_{3/2}-$ $^21_{3/2}$ & $\frac{\sqrt{N_c}}{\sqrt{2}(N_c+3)}$ & \\
$^48_{1/2}-$ $^21_{1/2}$ & $\frac{1}{2\sqrt{N_c+3}}$ & \\
$^48_{3/2}-$ $^21_{3/2}$ & $\frac{1}{2}\sqrt{\frac{5}{2(N_c+3)}}$ & \\
$^210_{1/2}-$ $^21_{1/2}$ & 0 & \\
$^210_{3/2}-$ $^21_{3/2}$ & 0 & \\
\hline
\end{tabular}}
\end{center}
\end{table}
%%%%%%%%%%%%%%%%%%%%%%%%%%%%%%%%%%%%%%%%%%%%%%%%%%%%%%%%%%%%%%%%%

%%%%%%%%%%%%%%%%%%%%%%%%%%%%%%%%%%%%%%%%%%%%%%%%%%%%%%%%%%%%%%%%%%%%%%%%%%%%
\vspace{2cm} 
 
{\bf Acknowledgments}
 The work of one of us (N. M.) was supported by F.R.S.-FNRS (Belgium).

%%%%%%%%%%%%%%%%%%%%%%%%%%%%%%%%%%%%%%%%%%%%%%%%%%%%%%%%%%%%%%%%%%%%%%%

%%%%%%%%%%%%%%%%%%%%%%%%%%%%%%%%%%%%%%%%%%%%%%%%%%%%%%%%%%%%%%%%%%%%%%%%%%

%%%%%%%%%%%%%%%%%%%%%%%%%%%%%%%%%%%%%%%%%%%%%%%%%%%%%%%%%%%%%%%%%%%%%%

\begin{thebibliography}{11}
\bibitem{HOOFT} G. 't Hooft, Nucl. Phys. {\bf 72} (1974) 461.

\bibitem{WITTEN} E. Witten, Nucl. Phys. {\bf B160} (1979) 57.

%\cite{Gervais:1983wq}
\bibitem{Gervais:1983wq}
  J.~L.~Gervais and B.~Sakita,
%%%   ``Large N QCD Baryon Dynamics: 
%%%     Exact Results From Its Relation To The Static
  %Strong Coupling Theory,''
  Phys.\ Rev.\ Lett.\  {\bf 52} (1984) 87;
  %%CITATION = PRLTA,52,87;%%
%\cite{Gervais:1984rc}
%%%\bibitem{Gervais:1984rc}
%%%  J.~L.~Gervais and B.~Sakita,
  %``Large-N baryonic soliton and quarks,''
  Phys.\ Rev.\ D {\bf 30} (1984) 1795.
  %%CITATION = PHRVA,D30,1795;%%



\bibitem{DM} R. Dashen and A. V. Manohar, Phys. Lett. {\bf B315} (1993) 425;
ibid {\bf B315} (1993) 438.

\bibitem{Jenk1}
E.~Jenkins, Phys.~Lett.  {\bf B315} (1993) 441;  {\bf B315} (1993) 431;  {\bf B315} (1993) 438.
 
\bibitem{DJM94}
R.~Dashen, E.~Jenkins, and A.~V.~Manohar,
Phys.~Rev.  {\bf D49} (1994) 4713.

\bibitem{DJM95}
R. Dashen, E. Jenkins, and A. V. Manohar,
Phys. Rev. {\bf D51} (1995) 3697.

\bibitem{CGO94}
C.~D.~Carone, H.~Georgi and S.~Osofsky,
Phys.~Lett. {\bf B322} (1994) 227.\\
M.~A.~Luty and J.~March-Russell,
Nucl.~Phys. {\bf B426} (1994) 71.\\
M.~A.~Luty, J.~March-Russell and  M.~White,
Phys.~Rev.  {\bf D51} (1995)  2332.



\bibitem{JL95}
E.~Jenkins and R.~F.~Lebed,
Phys.~Rev.  {\bf D52} (1995) 282.

\bibitem{DDJM96}
J.~Dai, R.~Dashen, E.~Jenkins, and A.~V.~Manohar,
Phys.~Rev.   {\bf D53} (1996) 273.

\bibitem{Jenkins:2009wv}
  E.~E.~Jenkins, A.~V.~Manohar, J.~W.~Negele and A.~Walker-Loud,
  %``A Lattice Test of 1/N_c Baryon Mass Relations,''
  Phys.\ Rev.\  D {\bf 81} (2010) 014502
%  [arXiv:0907.0529 [hep-lat]].
  %%CITATION = PHRVA,D81,014502;%%



\bibitem{CGKM} C. D. Carone, H. Georgi, L. Kaplan and D. Morin,
 Phys. Rev. {\bf D50} (1994) 5793.

\bibitem{Goi97}
J.~L. Goity, Phys. Lett.  {\bf B414} (1997) 140.

\bibitem{PY1} D. Pirjol and T.~M. Yan,
 Phys. Rev.  {\bf D57}  (1998) 1449.

\bibitem{PY2}  D. Pirjol and T.~M. Yan,  Phys. Rev. {\bf D57} (1998) 5434.

\bibitem{CCGL}
C.~E. Carlson, C.~D. Carone, J.~L. Goity and R.~F. Lebed,
 Phys. Lett. {\bf B438}  (1998) 327;
 Phys. Rev. {\bf D59}  (1999) 114008.


\bibitem{CaCa98}
C.~E.~Carlson and C.~D.~Carone,
 Phys.\ Lett.\  {\bf  B441} (1998) 363;
 Phys.\ Rev.\  {\bf  D58} (1998) 053005.


\bibitem{Pirjol:2003ye}
D.~Pirjol and C.~Schat, Phys. Rev. {\bf  D67} (2003) 096009.


%\cite{Cohen:2003tb}
\bibitem{Cohen:2003tb}
  T.~D.~Cohen and R.~F.~Lebed,
  %``New relations for excited baryons in large N(c) QCD,''
  Phys.\ Rev.\ Lett.\  {\bf 91}, 012001 (2003);
%  [arXiv:hep-ph/0301167].
  %%CITATION = PRLTA,91,012001;%%
   Phys.\ Rev.\ {\bf  D67}, 096008 (2003);

%\bibitem{COLEB}
%T. D. Cohen and R. F. Lebed, Phys. Rev. {\bf D67} (2003) 096008.



\bibitem{SGS}
C.~L. Schat, J.~L. Goity and N.~N. Scoccola,  Phys. Rev. Lett.
{\bf 88} (2002) 102002;
J.~L.~Goity, C.~L.~Schat and N.~N.~Scoccola,
 Phys.\ Rev.\  {\bf  D66} (2002) 114014.
 



%\cite{Matagne:2006zf}
%\bibitem{Matagne:2006zf}
%  N.~Matagne and F.~Stancu,
  %``Masses of (70,l+) baryons in the 1/N(c) expansion,''
%  Phys. Rev. {\bf D74} (2006) 034014.
%  [arXiv:hep-ph/0604122].
  %%CITATION = HEP-PH 0604122;%%


%\bibitem{MS1} N. Matagne and Fl. Stancu,  Phys. Rev. {\bf D71} (2005) 014010.

\bibitem{Matagne:2006dj}
  N.~Matagne and F.~Stancu,
  %``New look at the (70,1-) baryon multiplet in the 1/N(c) expansion,''
  Nucl.\ Phys.\  A {\bf 811} (2008) 291
  [arXiv:hep-ph/0610099].
  %%CITATION = NUPHA,A811,291;%%

\bibitem{HP} K. T. Hecht and S. C. Pang, J. Math. Phys. {\bf 10} (1969) 1571.


%\cite{Matagne:2006xx}
\bibitem{Matagne:2006xx}
  N.~Matagne and F.~Stancu,
  %``Matrix elements of SU(6) generators for baryons at arbitrary N(c),''
  Phys. Rev.  {\bf D73} (2006) 114025.
%  [arXiv:hep-ph/0603032].
  %%CITATION = HEP-PH 0603032;%%

\bibitem{Matagne:2008kb}
  N.~Matagne and F.~Stancu,
  %``Matrix elements of SU(6) generators for baryons with arbitrary $N_c$ quarks
  %in mixed symmetric states $[N_c-1,1]$,''
  Nucl.\ Phys.\  A {\bf 826} (2009) 161.
%  [arXiv:0812.1365 [hep-ph]].
  %%CITATION = NUPHA,A826,161;%%
  
  
%\bibitem{book} Fl. Stancu, {\it Group Theory in Subnuclear Physics},
%Clarendon Press, Oxford, (1996) Ch. 4.

\bibitem{Stancu:1991rc}
  F.~Stancu,
  ``Group theory in subnuclear physics,''
  Oxford Stud.\ Nucl.\ Phys.\  {\bf 19} (1996) 1.
  %%CITATION = 00308,19,1;%%
  
\bibitem{HECHT} K. T. Hecht, Nucl. Phys. {\bf 62}, 1 (1965).  
  
 
\bibitem{DESWART} J. J. De Swart, Rev. Mod. Phys. {\bf 35}, 916 (1963).
  
  
\bibitem{Matagne:2005gd}
  N.~Matagne and F.~Stancu,
  %``Excited (70,l+) baryons in large N(c) QCD,''
  Phys.\ Lett.\  B {\bf 631} (2005) 7.
%  [arXiv:hep-ph/0505118].
  %%CITATION = PHLTA,B631,7;%%
  
  
\bibitem{Scoccola:2007sn}
  N.~N.~Scoccola, J.~L.~Goity and N.~Matagne,
  %``Analysis of Negative Parity Baryon Photoproduction Amplitudes in the
  %$1/N_c$ Expansion,''
  Phys.\ Lett.\  B {\bf 663} (2008) 222.
%  [arXiv:0711.4203 [hep-ph]].
  %%CITATION = PHLTA,B663,222;%%
  
  
\bibitem{BLED}
 N. Matagne and Fl. Stancu,
Proceedings of the Bled Workshop on {\it Dressing hadrons},
Bled, Slovenia, July 4-11, 2010,
Bled Workshops in Physics, vol. 11 no. 1:  37-43 , 2010,
eds. B. Golli, M. Rosina, S. Sirca,
arXiv:1101.3845 [hep-ph].

\bibitem{PDG}
K. Nakamura {\it et al.} [Particle Data Group], J. Phys. {\bf G37} (2010) 075021.
 
 \bibitem{Hey:1974nc}
  A.~J.~G.~Hey, P.~J.~Litchfield and R.~J.~Cashmore,
  %``SU(6)-W And Decays Of Baryon Resonances,''
  Nucl.\ Phys.\  B {\bf 95} (1975) 516.
  %%CITATION = NUPHA,B95,516;%%

\bibitem{Capstick:1986bm}
  S.~Capstick and N.~Isgur,
  %``Baryons In A Relativized Quark Model With Chromodynamics,''
  Phys.\ Rev.\  D {\bf 34} (1986) 2809.
  %%CITATION = PHRVA,D34,2809;%%

 
\bibitem{Glozman:1997ag}
  L.~Y.~Glozman, W.~Plessas, K.~Varga and R.~F.~Wagenbrunn,
  %``Unified description of light- and strange-baryon spectra,''
  Phys.\ Rev.\  D {\bf 58} (1998) 094030.
%  [arXiv:hep-ph/9706507].
  %%CITATION = PHRVA,D58,094030;%%

\bibitem{Melde:2008yr}
  T.~Melde, W.~Plessas, B.~Sengl,
  %``Quark-Model Identification of Baryon Ground and Resonant States,''
  Phys.\ Rev.\  {\bf D77 } (2008)  114002.
 % [arXiv:0806.1454 [hep-ph]].
 
\bibitem{Matagne:2010se}
  N.~Matagne and F.~Stancu,
  %``New look at the $[{\bf 70},1^-]$ nonstrange and strange baryons in the
  %$1/N_c$ expansion,''
  arXiv:1011.5758 [hep-ph].
  %%CITATION = ARXIV:1011.5758;%%

 
 
  
 \bibitem{Pakvasa:1999zv}
  S.~Pakvasa and S.~F.~Tuan,
  %``Lambda/s(1405) and negative parity baryon states,''
  Phys.\ Lett.\  B {\bf 459} (1999) 301.
%  [arXiv:hep-ph/9903551].
  %%CITATION = PHLTA,B459,301;%%

\bibitem{COLEB2}
T. D. Cohen and R. F. Lebed, Phys. Rev. {\bf D72} (2005) 056001.

 
\bibitem{GarciaRecio:2006wb}
  C.~Garcia-Recio, J.~Nieves and L.~L.~Salcedo,
  %``Large N(c) Weinberg-Tomozawa interaction and negative parity s-wave  baryon
  %resonances,''
  Phys.\ Rev.\  D {\bf 74} (2006) 036004.
%  [arXiv:hep-ph/0605059].
  %%CITATION = PHRVA,D74,036004;%%
 
\bibitem{Hyodo:2007np}
  T.~Hyodo, D.~Jido and L.~Roca,
  %``Structure of the (1405) baryon resonance from its large N(c) behavior,''
  Phys.\ Rev.\  D {\bf 77} (2008) 056010.
%  [arXiv:0712.3347 [hep-ph]].
  %%CITATION = PHRVA,D77,056010;%%

\bibitem{MattisMukerjee} M. P. Mattis and M. Mukerjee,  
Phys. Rev. Lett.  {\bf 61} (1988) 1344;  M. P. Mattis, Phys. Rev. Lett. 
{\bf 63} (1989) 1455.

\bibitem{COLEB1}
T. D. Cohen and R. F. Lebed, Phys. Rev. {\bf D67} (2003) 096008.


\end{thebibliography}
\end{document}